\documentclass[aps,prd,preprint,amsmath,amssymb,amsfonts,showpacs]{revtex4-1}

\usepackage{amsthm}
\usepackage{graphicx}

\newtheorem{theorem}{Theorem}

\newcommand{\bbR}{\mathbb{R}}
\newcommand{\bbZ}{\mathbb{Z}}
\newcommand{\scri}{\mathcal{I}}
\newcommand{\calA}{\mathcal{A}}
\newcommand{\calD}{\mathcal{D}}
\newcommand{\calO}{\mathcal{O}}
\newcommand{\calP}{\mathcal{P}}

\newcommand{\del}{\partial}

\newcommand{\sign}{\operatorname{sign}}
\renewcommand{\Re}{\operatorname{Re}}
\renewcommand{\Im}{\operatorname{Im}}

\newcommand{\overbar}[1]{\mkern 1.5mu\overline{\mkern-1.5mu#1\mkern-1.5mu}\mkern 1.5mu}

\begin{document}

%\raggedbottom

\title{Holography and quantum states in elliptic de Sitter space}

\author{Illan F. Halpern}
\email{illan@berkeley.edu}
\affiliation{Department of Physics, University of California, Berkeley, CA, 94720, USA}
\author{Yasha Neiman}
\email{yashula@gmail.com}
\affiliation{Perimeter Institute for Theoretical Physics, 31 Caroline Street N, Waterloo, ON, N2L 2Y5, Canada}

\date{\today}

\begin{abstract}
We outline a program for interpreting the higher-spin dS/CFT model in terms of physics in the causal patch of a dS observer. The proposal is formulated in ``elliptic'' de Sitter space $dS_4/\bbZ_2$, obtained by identifying antipodal points in $dS_4$. We discuss recent evidence that the higher-spin model is especially well-suited for this, since the antipodal symmetry of bulk solutions has a simple encoding on the boundary. For context, we test some other (free and interacting) theories for the same property. Next, we analyze the notion of quantum field states in the non-time-orientable $dS_4/\bbZ_2$. We compare the physics seen by different observers, with the outcome depending on whether they share an arrow of time. Finally, we implement the marriage between higher-spin holography and observers in $dS_4/\bbZ_2$, in the limit of free bulk fields. We succeed in deriving an observer's operator algebra and Hamiltonian from the CFT, but not her S-matrix. We speculate on the extension of this to interacting higher-spin theory.
\end{abstract}


\maketitle
\tableofcontents
\newpage

\section{Introduction}

\subsection{dS/CFT and observers}

It could once be said that quantum gravity is the great open problem in theoretical physics. With advances in string theory and particularly AdS/CFT \cite{Aharony:1999ti,Witten:1998qj}, this statement must now be qualified. Quantum gravity appears to be understood in principle, as long as we restrict to questions posed at spatial infinity in a world with negative cosmological constant $\Lambda<0$. The \emph{remaining} conceptual problem is what to make of quantum gravity in finite regions, in particular for observers who are confined inside a causal horizon. This problem is brought into focus by the evidence that our Universe has a positive cosmological constant $\Lambda>0$, and that therefore \emph{every} observer is surrounded by a cosmological horizon of bounded area.

The most natural playground for quantum gravity in finite regions is de Sitter space (dS) -- the maximally symmetric spacetime that contains horizons and has $\Lambda>0$. Cosmological horizons like the ones in dS are simpler than black hole horizons: they do not come with singularities, making it possible to disentangle the two issues. Finally, dS can be viewed as an analytical continuation of anti-de Sitter (AdS), creating a possibility to import successful techniques from the $\Lambda<0$ case. This line of thinking leads to dS/CFT \cite{Strominger:2001pn} -- holography in de Sitter space. Throughout the paper, we will specialize to the physically relevant spacetime dimension, i.e. $dS_4$.

dS/CFT is essentially an attempt to cheat: one takes a system that should differ conceptually from AdS due to its different causal structure, and tries nevertheless to use the same toolkit. The price to pay is that the physical interpretation of the AdS/CFT dictionary must change, and the result may end up not telling us what we wanted to know about de Sitter physics. There are two reasons for this. First, the boundary of dS is not a place, but a pair of times: the infinite past $\scri^-$ and the infinite future $\scri^+$ (each with the geometry of a conformal 3-sphere $S_3$). Second, the fields on either of the $\scri^\pm$ causally determine all of spacetime, while the true object of interest is the physics inside an observer's causal patch.

To be more specific, the leading paradigm for dS/CFT \cite{Maldacena:2002vr,Harlow:2011ke} equates the CFT partition function with the \emph{Hartle-Hawking wavefunction} \cite{Hartle:1983ai} of the bulk gravitational theory, evaluated on the asymptotic future time slice $\scri^+$:
\begin{align}
 Z_{\text{CFT}}[\text{sources on }S_3] = \Psi_{\text{HH}}[\text{fields on }\scri^+] \ . \label{eq:HH}
\end{align}
This construction implies a Hilbert space of global bulk states, constructed trivially using fields on $\scri^+$ as a configuration basis. The CFT's job is then to pick a special state out of this Hilbert space. We are left with no notion of time evolution, or of the state space in an observer's causal patch. For instance, it's unclear how to determine whether the state space accessible to an observer is finite-dimensional, as suggested by the horizon entropy formula \cite{Gibbons:1977mu,Bekenstein:1980jp,Bousso:1999xy}. 

On the upside, there is a proposed concrete realization \cite{Anninos:2011ui} of the duality \eqref{eq:HH}. The bulk ``gravity'' in this realization is Vasiliev's bosonic higher-spin theory \cite{Vasiliev:1995dn,Didenko:2014dwa}, while the CFT is a vector model. This proposal analytically continues the higher-spin AdS/CFT duality \cite{Klebanov:2002ja,Sezgin:2003pt,Giombi:2012ms} from negative to positive $\Lambda$ -- a step that appears impossible for the more standard, string-based holographic models.

The broad goal of this paper is to approach the problem of quantum gravity in observers' causal patches from within the dS/CFT model of \cite{Anninos:2011ui}. 
Thus, we aim to bring the ``cheating'' in dS/CFT to fruition: after importing from AdS the tools for describing quantities at infinity, we wish to further translate their output into statements about states and evolution in an observer's causal patch. This will require re-interpreting the bulk side of the duality \eqref{eq:HH}.

The key ingredient in our proposal is to replace the bulk spacetime $dS_4$ with Schrodinger's ``elliptic'' de Sitter space $dS_4/\bbZ_2$, obtained by identifying antipodal points in $dS_4$ \cite{Folacci:1986gr}. This was first proposed in the dS/CFT context in \cite{Parikh:2002py,Parikh:2004ux,Parikh:2004wh}. This ``folding in half'' of the spacetime respects the de Sitter isometries. However, the spacetime loses its time-orientability; in particular, past infinity $\scri^-$ becomes identified with future infinity $\scri^+$. This non-orientability is a strictly global property: there are no closed timelike curves, and each observer's causal patch remains time-orientable. 

The main upshot of the switch to $dS_4/\bbZ_2$ is that an observer's causal patch (now identified with its antipode) causally spans the entire spacetime, including the asymptotic boundary $\scri^-\equiv\scri^+\equiv\scri$. This creates a possibility to achieve the desired translation between the CFT at $\scri$ and the causal patch. As discussed below, we successfully carry out this translation in the limit of free bulk fields in the higher-spin model. Specifically, we find that the operator algebra and the Hamiltonian in the causal patch can be read off from the CFT partition function.

\subsection{Structure of the paper and summary of results}

The paper's structure and main results are as follows. In section \ref{sec:geometry}, we review the spacetime geometry of $dS_4$ and $dS_4/\bbZ_2$, with an emphasis on the concepts of observers and operationally observable regions. In section \ref{sec:encoding}, we argue that the higher-spin model of \cite{Anninos:2011ui} is especially well-suited to the antipodal identification proposal of \cite{Parikh:2002py}. This is due to observations \cite{Neiman:2014npa} by one of the authors, which suggest that in higher-spin gravity, antipodal symmetry of the bulk solution is encoded (to all orders in the interaction) by a simple condition at $\scri$: the vanishing of one of the two types of boundary data for each bulk field. For context, we test more conventional field theories in $dS_4$ for this property. At the linearized level, we find that it holds for free massless fields of all spins, as well as for the ``partially massless'' fields of \cite{Higuchi:1986wu,Deser:2001us,Zinoviev:2001dt}, but not for massive fields. Beyond free fields, the property holds at 3-point level in Yang-Mills and General Relativity (GR), but fails for interacting scalars. It appears, therefore, that this simple boundary encoding of the $dS_4/\bbZ_2$ bulk topology is associated with masslessness and gauge symmetry, which find their full expression in higher-spin theory.

In section \ref{sec:QFT}, we turn to analyze the basic concepts of quantum field theory in $dS_4/\bbZ_2$, clarifying and extending the treatment in \cite{Hackl:2014txa}. The challenge is to make sense of quantum states and operators in a non-time-orientable spacetime. As noted in \cite{Hackl:2014txa}, a quantum theory in $dS_4/\bbZ_2$ can only be formulated after choosing an observer. One can then ask how the worldviews of different observers relate to each other, and what is the global structure that underlies them. In \cite{Hackl:2014txa}, a recipe was proposed for translating between observers. It was pointed out that different observers under this recipe may not agree on operator expectation values. Here, we will clarify the recipe of \cite{Hackl:2014txa}, demonstrating in particular that observers \emph{do} agree in regions where they perceive the same arrow of time. The underlying structure behind the observers' worldviews is identified as the space of antipodally symmetric states in ordinary $dS_4$.

In section \ref{sec:dS/CFT}, we use the insights from the previous sections to make contact with dS/CFT. We consider the higher-spin model of \cite{Anninos:2011ui} at the level of 2-point functions, i.e. the free-field limit in the bulk. In this limit, we find that \emph{the CFT partition function encodes the operator algebra and Hamiltonian in an observer's causal patch} in terms of fields on $\scri$. As mentioned above, this is made possible by the switch to $dS_4/\bbZ_2$, since the fields in the causal patch are now causally equivalent to those on $\scri$. On the other hand, we find that the ``S-matrix'' of transition amplitudes between the observer's past and future horizons is \emph{not} encoded similarly by the CFT. The reason is that the S-matrix is defined by certain complex phases, which are not captured by the real CFT partition function. Finally, in section \ref{sec:discuss}, we discuss the challenges of upgrading our approach from free bulk fields to full-fledged higher-spin gravity.

\section{Geometry and observers in ordinary and elliptic de Sitter space} \label{sec:geometry}

\subsection{De Sitter space embedded in $\bbR^{1,4}$} \label{sec:geometry:embedding}

We define de Sitter space $dS_4$ as the hyperboloid of unit spacelike radius in (4+1)d flat spacetime: 
\begin{align}
 dS_4 = \left\{ x\in\bbR^{1,4}\, |\ x_\mu x^\mu = 1 \right\} \ . \label{eq:dS}
\end{align}
This implies a choice of units where the cosmological constant is $\Lambda = 3$. The vector indices $(\mu,\nu,\dots)$ in $\bbR^{1,4}$ are raised and lowered by the flat metric $\eta_{\mu\nu}$ with signature $(-,+,+,+,+)$. We use the same indices for vectors $v^\mu$ in the tangent space of $dS_4$ at a point $x$, with the understanding that $x_\mu v^\mu = 0$. The projector from $\bbR^{1,4}$ into the tangent space at $x\in dS_4$ reads:
\begin{align}
 P_\mu^\nu(x) = \delta_\mu^\nu - x_\mu x^\nu \ . \label{eq:projector}
\end{align}
The lowered-index version $P_{\mu\nu}(x) = \eta_{\mu\nu} - x_\mu x_\nu$ defines the metric of $dS_4$. The covariant derivative in $dS_4$ is just the flat derivative in $\bbR^{1,4}$, projected back into the tangent space:
\begin{align}
 \nabla_\mu v_\nu = P_\mu^\rho(x) P_\nu^\sigma(x) \del_\rho v_\sigma \ . \label{eq:nabla}
\end{align}
The commutator of covariant derivatives reads:
\begin{align}
 [\nabla_\mu,\nabla_\nu]v^\rho = 2\delta_{[\mu}^\rho v_{\nu]}^{\vphantom{\rho}} \ . \label{eq:commutator}
\end{align}
The de Sitter isometry group $O(1,4)$ is just the group of rotations and reflections in $\bbR^{1,4}$.

The boundaries of $dS_4$ may be equated with the asymptotes of the hyperboloid \eqref{eq:dS}. These are the spaces of past-pointing and future-pointing null directions in $\bbR^{1,4}$, which we identify respectively as past infinity $\scri^-$ and future infinity $\scri^+$. Each of the two boundary components $\scri^\pm$ has the geometry of a spacelike conformal 3-sphere $S_3$. Points on $\scri^\pm$ are represented by null vectors $\ell^\mu\in\bbR^{1,4}$, defined up to rescalings $\ell^\mu\rightarrow \alpha\ell^\mu$. These local rescalings can be identified with Weyl transformations on $\scri^\pm$. Quantities on $\scri^\pm$ with conformal weight $\Delta$ can be represented as functions $f(\ell)$ on the lightcone that scale as $f\rightarrow \alpha^{-\Delta}f$. The time-orientation-preserving component $O^\uparrow(1,4)$ of the de Sitter group acts on $\scri^\pm$ as the conformal group of the 3-sphere.

A bulk point $x\in dS_4$ is said to approach the boundary point $l\in\scri^\pm$ if the vector components of $x^\mu$ behave as:
\begin{align}
 x^\mu \rightarrow \frac{l^\mu}{z} \ , \quad \text{with} \ z\rightarrow 0 \ . \label{eq:asymptote}
\end{align}
This implies that the radius-vector $x^\mu$ is highly boosted in some fixed frame. The frame dependence is unavoidable, since the statement that a bulk point is ``close to infinity'' isn't invariant under large translations. Scalar fields in the bulk that scale as $f\sim z^\Delta$ in the asymptotic limit \eqref{eq:asymptote} become fields with conformal weight $\Delta$ on $\scri^\pm$. 

Three-dimensional vectors $v^\mu$ tangent to $\scri^\pm$ at a point $l$ can again be written using (4+1)d indices, with the understanding that they're tangent to the lightcone in $\bbR^{1,4}$, i.e. $l_\mu v^\mu = 0$, and defined up to shifts $v^\mu\rightarrow v^\mu + \alpha l^\mu$. The scaling rules become a bit subtle, since a tangent vector on $\scri^\pm$ has a divergent length $\sim 1/z$ in the limiting procedure \eqref{eq:asymptote}. Thus, a bulk field $f^{\mu_1\dots\mu_k}{}_{\nu_1\dots\nu_n}$ with indices tangent to $\scri^\pm$ in the limit \eqref{eq:asymptote} is said to have conformal weight $\Delta$ if its components scale as $f^{\mu_1\dots\mu_k}{}_{\nu_1\dots\nu_n}\sim z^{\Delta-k+n}$ in an orthonormal bulk basis.

\subsection{Horizons and observers in $dS_4$} \label{sec:geometry:observer}

A boundary point $\ell\in\scri^\pm$ casts a lightcone into the $dS_4$ bulk. This is the cosmological horizon associated with the boundary point $\ell$. It is a 3d null hypersurface, given by the intersection of the $dS_4$ hyperboloid \eqref{eq:dS} with the null hyperplane $\ell_\mu x^\mu = 0$. Its spatial sections are unit 2-spheres. The boundary point antipodal to $\ell$, defined by the null vector $-\ell^\mu$, generates the same horizon. In other words, the lightcone of a point on $\scri^-$ refocuses at the antipodal point on $\scri^+$. 

\begin{figure}%
\centering%
\includegraphics[scale=1]{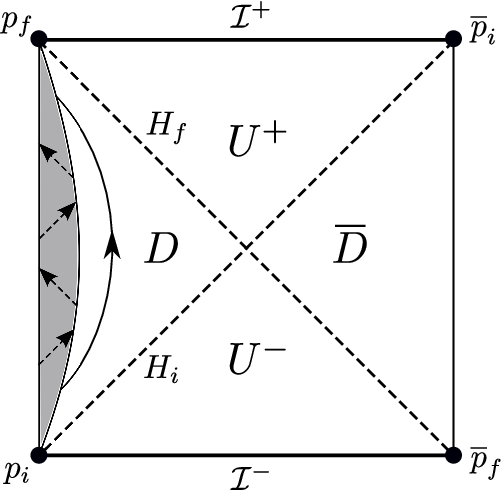} \\
\caption{A Penrose diagram of $dS_4$. Past (future) infinity is denoted by $\scri^-$ ($\scri^+$). The boundary points $p_i\in\scri^-$ and $p_f\in\scri^+$ define horizons $H_i,H_f$ that divide the spacetime into quadrants: the causal patch $D$, the antipodal causal patch $\overbar D$ and the quadrants $U^\pm$ containing $\scri^\pm$. The antipodes of $p_i,p_f$ are denoted by $\bar p_i,\bar p_f$. The shaded area inside the causal patch depicts an extended observer whose constituent parts exchange causal signals, depicted as dashed arrows. The worldline to the right depicts a probe launched by the observer, which eventually returns with information.}
\label{fig:observer} 
\end{figure}

We identify an \emph{observer} in $dS_4$ with a pair of (non-antipodal) boundary points $p_i\in\scri^-$ and $p_f\in\scri^+$. Any such pair of points is equivalent to any other by an $SO^\uparrow(1,4)$ transformation. The points $p_i,p_f$ can be loosely thought of as the past and future endpoints of the observer's worldline. Their respective lightcones $H_i,H_f$ are the observer's past and future horizons. The two horizons intersect at a unit 2-sphere, known as the bifurcation surface. Together, they divide de Sitter space into four quadrants: the observer's causal patch, the antipodal causal patch, a quadrant containing $\scri^-$ and another containing $\scri^+$. See figure \ref{fig:observer}. 

For future reference, the metric of the observer's causal patch in static coordinates reads:
\begin{align}
 ds^2 = -\cos^2\chi\,dt^2 + d\chi^2 + \sin^2\chi\left(d\theta^2 + \sin^2\theta\,d\phi^2\right) \ . \label{eq:static}
\end{align}
The subgroup of $O(1,4)$ isometries that preserves the choice of endpoints $p_i,p_f$ is $\bbR\times O(3)$. This will be the symmetry group of the Hilbert space of states accessible to the observer. The $\bbR$ factor refers to translations in the time coordinate $t$, while the $O(3)$ refers to rotations/reflections of the $(\theta,\phi)$ 2-spheres.

Let us now refine our physical picture of an observer. As a first step, we can imagine her as a pointlike particle, following a timelike worldline with asymptotic endpoints $p_i,p_f$. This worldline may or may not be the geodesic that stretches from $p_i$ to $p_f$. Regardless of the worldline's detailed shape, the spacetime region into which the observer can (at some time or another) send causal signals is the half of $dS_4$ to the future of $H_i$. Similarly, the region from which the observer can receive signals is the half-spacetime to the past of $H_f$. The intersection of these two regions is the observer's causal patch, also known as the static patch or the causal diamond.

The above standard argument justifies our treatment of the endpoints $p_i,p_f$ as the only relevant parameters defining an observer, and establishes the roles of $H_i,H_f$ as the past and future horizons. However, this picture involves two idealizations, which we will now address. First, an observer with the ability to store and manipulate information cannot be pointlike (due to information bounds such as Bekenstein's, if nothing else). It is better to imagine the observer as an extended object, consisting of components with their own individual worldlines. However, to function as a single entity, these components must exchange causal signals, which is only possible if they stay inside each other's horizons. For the components to exchange signals indefinitely, their worldlines must share the same two boundary points $p_i,p_f$. Thus, our ``fattened'' observer is still confined to the causal patch defined by the endpoints $p_i,p_f$ and their associated horizons $H_i,H_f$. Figure \ref{fig:observer} depicts this situation. We note the similarity with the notion of isolated systems in de Sitter space \cite{Ashtekar:2014zfa}.

For an observer with a bounded lifespan, the spacetime region spanned by the components' worldlines \emph{while} they maintain causal contact is still contained inside a causal patch (say, the one defined by extending one of the worldlines into the asymptotic past and future). In this sense, we do not lose generality by considering eternal observers, which we will do from now on.

The second idealization in the standard picture is that the observer can ``see'' the entire contents of her past lightcone, including e.g. $\scri^-$. Strictly speaking, we never see anything outside our eyes: an observer can only interact locally with objects that coincide with her in spacetime. Any conclusions about the outside world must involve assumptions about the dynamics. For ordinary eyesight, such assumptions may include e.g. the optical properties of the medium. The observer may theoretically deduce e.g. field values in spacetime regions that aren't contiguous with her (provided sufficient initial data), but she cannot \emph{observe} them experimentally. As a possible exception to this statement, one can imagine the observer launching a probe, which later returns with data about distant regions. However, having accepted that the observer is a composite extended object, we can simply incorporate such probes into our definition of her -- a generalized limb, so to speak. In particular, for the probe to return to the observer with its collected information, it must remain inside her causal patch, just like any other ``body part''.

To conclude:
\begin{enumerate}
 \item An observer in $dS_4$, including any probes that she may launch, can be thought of as a bundle of wordlines, all staying within each other's causal horizons. 
 \item Only spacetime regions contiguous with the observer are truly observable, as opposed to the entirety of her past lightcone. 
 \item The largest possible observable region, i.e. the largest region that can be physically spanned by an observer and her probes, is the causal patch defined by two boundary points $p_i\in\scri^-$ and $p_f\in\scri^+$. For the purposes of this paper, we identify such a patch with a choice of observer.
\end{enumerate}

\subsection{Antipodal identification and the geometry of $dS_4/\bbZ_2$} \label{sec:geometry:antipodal}

Every point in $dS_4$ has a point antipodal to it in space and time. In the embedding-space picture \eqref{eq:dS}, this antipodal map is simply $x^\mu\rightarrow -x^\mu$. It is invariant under the $dS_4$ isometry group $O(1,4)$, and is in fact its unique non-trivial central element. In the standard CPT classification of discrete symmetries, the antipodal map is of the CT type \cite{Neiman:2014npa}. On the square Penrose diagram of $dS_4$, it is represented by a reflection of both axes around the center. However, one should remember that each point on the Penrose diagram is really a 2-sphere, and the antipodal map also acts within these 2-spheres. In particular, the separation between antipodal points is always spacelike, even though it may appear timelike on the Penrose diagram. When extended to the boundary of $dS_4$, the antipodal map relates a point $\ell^\mu\in\scri^+$ to its image $-\ell^\mu\in\scri^-$. Since the lightcone of $-\ell^\mu$ refocuses at $\ell^\mu$, one can say that antipodes on $\scri^\pm$ are null-separated, with an infinite affine distance between them.

For tensors on $dS_4$, we define the action of the antipodal map as the appropriate pullback/push-forward. In the embedding-space notation of section \ref{sec:geometry:embedding}, this becomes $f_{\mu_1\dots\mu_k}(x) \rightarrow (-1)^k f_{\mu_1\dots\mu_k}(-x)$. With this convention, the $dS_4$ covariant derivative $\nabla_\mu$ is antipodally even, while the (3+1)d Levi-Civita tensor $\epsilon^{\mu\nu\rho\sigma} = x_\lambda\epsilon^{\lambda\mu\nu\rho\sigma}$ is antipodally odd.

Elliptic de Sitter space $dS_4/\bbZ_2$ is the quotient space obtained by topologically identifying antipodal points in $dS_4$. Like $dS_4$, it is a maximally symmetric spacetime. Its isometry group is $O(1,4)/\bbZ_2 = SO(1,4)$, where the $\bbZ_2$ is generated by the antipodal map. The boundary of $dS_4/\bbZ_2$ is a single conformal 3-sphere $\scri$, which results from the identification of $\scri^-$ and $\scri^+$. As a manifold, $dS_4/\bbZ_2$ is non-orientable and doubly-connected. In addition, its metric is non-orientable in time, which will be important below.

In practice, it's convenient to imagine fields on $dS_4/\bbZ_2$ as antipodally symmetric fields on ordinary $dS_4$. A path connecting two antipodal points in $dS_4$ becomes an incontractible loop in $dS_4/\bbZ_2$. Antipodally even tensors on $dS_4$ take values in the trivial tensor bundle on $dS_4/\bbZ_2$, while antipodally odd tensors take values in a ``twisted'' bundle, where the fiber changes sign upon traversing an incontractible loop. The Levi-Civita tensor $\epsilon^{\mu\nu\rho\sigma}$ can be viewed as belonging to this odd bundle. Since the antipodal map is of the CT type, $dS_4/\bbZ_2$ can only support field theories that preserve CT, or, equivalently, P. For consistent evolution on $dS_4/\bbZ_2$, fields that transform under parity as tensors/pseudotensors must be antipodally even/odd, respectively.

\subsection{Causal structure and observers in $dS_4/\bbZ_2$} \label{sec:geometry:elliptic_observers}

We now turn to the causal structure of $dS_4/\bbZ_2$. On one hand, this spacetime is non-time-orientable, which will play a crucial role in the discussion of quantum theory in section \ref{sec:QFT}. On the other hand, this non-orientability is global in nature, and cannot be detected by any observer. In particular, the antipodal identification doesn't generate closed timelike curves, since antipodal points are always spacelike-separated. 

For a more detailed picture, let's consider the horizon structure for an observer in $dS_4/\bbZ_2$, as we've done for a $dS_4$ observer in section \ref{sec:geometry:observer}. An observer is once again defined by a pair of boundary points $p_i$,$p_f$. Since $\scri^-$ and $\scri^+$ have been identified, these are now just two points on the single boundary 3-sphere $\scri$. The boundary points again define horizons $H_i$,$H_f$. The antipodal map in $dS_4$ maps each horizon to itself, so that the $dS_4/\bbZ_2$ horizons are folded-in-half versions of the ones in $dS_4$. Similarly, the bifurcation surface $H_i\cap H_f$ gets mapped to itself, becoming an antipodally-identified 2-sphere $S_2/\bbZ_2$ in $dS_4/\bbZ_2$.

\begin{figure}%
\centering%
\includegraphics[scale=1]{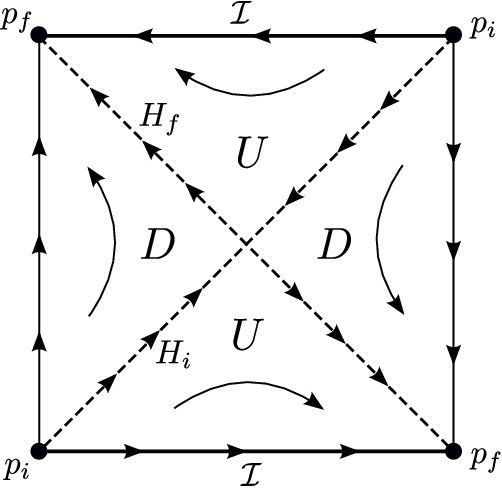} \\
\caption{A Penrose diagram of $dS_4/\bbZ_2$. Opposite points in the diagram are identified. Past and future infinity are identified into a single boundary $\scri$. The observer's worldline endpoints $p_i,p_f\in\scri$ define horizons $H_i,H_f$, which divide the spacetime into two independent ``quadrants'': the causal patch $D$ (identified with its antipode) and its complement $U$, which contains $\scri$. The arrows denote the direction of the Killing vector $\xi^\mu$ that generates time translations in the causal patch. The observer induces a time orientation in regions where $\xi^\mu$ is causal, i.e. in the causal patch and on the horizons.}
\label{fig:time} 
\end{figure}

The horizons $H_i$,$H_f$ divide spacetime into \emph{two} regions, instead of the four regions in figure \ref{fig:observer}. This is because the causal patch $D$ from figure \ref{fig:observer} is now identified with its antipode $\overbar D$, and the two quadrants $U^\pm$ are identified with each other. The observer can again be thought of as a (potentially space-filling) set of worldlines inside the causal patch, each stretching from $p_i$ to $p_f$. On the $dS_4$ Penrose diagram, one should think of the observer as also living in the antipodal causal patch, with worldlines going ``backwards in time'' from $\bar p_i$ to $\bar p_f$. This new situation is depicted in figure \ref{fig:time}.

Though we lack a global time orientation, we can still make a distinction between $p_i$ as the observer's \emph{initial} point and $p_f$ as her \emph{final} point. This defines the direction of the observer's proper time, and with it a natural time orientation in the causal patch $D$. This choice of time orientation can be extended to the horizons, but is discontinuous across the bifurcation surface. This is the best we can do: the embedding of e.g. an $S_3/\bbZ_2$ equatorial spatial slice into $dS_4/\bbZ_2$ is non-orientable, so one cannot make an everywhere continuous choice for the direction of the timelike normal.

The distinction between $p_i$ and $p_f$ does \emph{not} induce a time orientation in the spacetime region $U^-\equiv U^+\equiv U$ outside the causal patch. One could decide that the direction towards $\scri$ should count as ``future'' or ``past'', but there is no preference for either choice, and each would be discontinuous with the time orientation in the causal patch across one of the horizons $H_i$,$H_f$. 

The time orientation in the causal patch can be identified with the direction of the (antipodally even) Killing vector $\xi^\mu$, which generates time translations $\del_t$ in the static coordinates \eqref{eq:static}. In the embedding-space picture where the boundary points $p_i$,$p_f$ are viewed as null directions in $\bbR^{1,4}$, these time translations are just boosts in the $p_i\wedge p_f$ plane. The direction of $\xi^\mu$ in different regions of the Penrose diagram is illustrated in figure \ref{fig:time}. The lack of time orientation in the region $U$ outside the causal patch can be attributed to the fact that $\xi^\mu$ is spacelike there.

In $dS_4$, it is usually said that the observer can ``see'' the half-spacetime to the past of the future horizon $H_f$. With the antipodal identification, that would imply that the observer sees the \emph{entire} spacetime except $H_f$ itself. However, in keeping with the analysis of section \ref{sec:geometry:observer}, it is more correct to say that, as in ordinary $dS_4$, the observer can only see inside her causal patch. \emph{Unlike} in ordinary $dS_4$, the fields in the causal patch now (classically) determine the fields everywhere else, in particular at $\scri$. Indeed, in the $dS_4$ picture, antipodally symmetric Cauchy data in the causal patch \emph{together with its antipode} can be evolved into a unique antipodally symmetric solution in spacetime. This is our main motivation for studying $dS_4/\bbZ_2$, as it provides hope for linking holography at $\scri$ to physics in causal patches. It does not mean, however, that $\scri$ is operationally observable; only the causal patch is.

To conclude, $dS_4/\bbZ_2$ has no global time orientation. However, each observer defines a time orientation within her causal patch, induced by the ordering of her worldline endpoints $p_i$,$p_f$. Thus, the causal patch in $dS_4/\bbZ_2$ plays a dual role: it is both the observable spacetime region, and the one where the observer can assign a time orientation.

\section{Boundary encoding of the antipodal identification} \label{sec:encoding}

\subsection{Overview}

A bulk field in $dS_4$ has two types of asymptotic boundary data, corresponding to two different falloff rates, i.e. conformal weights at $\scri^\pm$. In this section, we discuss how in certain bulk theories, antipodal symmetry of a solution in $dS_4$ is equivalent to the \emph{vanishing} of one of these two types of boundary data. Such a simple boundary encoding of antipodal symmetry is a plausible prerequisite for describing physics in $dS_4/\bbZ_2$ using a boundary CFT. We will see its utility concretely in section \ref{sec:dS/CFT}, where we present a holographic treatment of free quantum fields in the bulk. In the present section, we consider classical bulk theories, both free and interacting.

In a \emph{free} bulk theory, the solution space can be decomposed into antipodally even and odd subspaces. In each subspace, there will be some linear (in general, non-local) relation between the two types of boundary data on $\scri$. We are interested in theories where this relation is just the vanishing of one of the two types of boundary data. In \cite{Ng:2012xp}, it was noticed that a conformally-coupled massless scalar has this property. In \cite{Neiman:2014npa}, the result was extended to massless gauge fields of all integer spins $s\geq 1$. In this section, we extend the analysis to other free theories, showing that:
\begin{enumerate}
 \item The property doesn't hold for massive fields.
 \item It holds for the ``partially massless'' gauge fields of \cite{Deser:2001us}.
 \item For the minimally-coupled massless scalar, it holds for antipodally even solutions, but not quite for antipodally odd ones.
\end{enumerate}
The overall implication is that the simple boundary encoding of antipodal symmetry is associated with masslessness. The special role of massless fields can be understood as a consequence of their lightlike propagation. Since the lightcone of a point on $\scri^-$ refocuses at its antipode on $\scri^+$, null propagation implies that data at the $\scri^-$ point is translated directly into data at its $\scri^+$ antipode. It is then natural for each type of boundary data to be associated with an (even or odd) antipodal symmetry.

At the interacting level, we recall that antipodally symmetric solutions only occur in parity-invariant theories, and then each field's antipodal symmetry sign is determined by its parity. Therefore, unlike in the free case, there will generally be just one set of antipodally symmetric solutions, instead of several subspaces with different sign choices. The question is then whether these antipodally symmetric solutions are associated with the vanishing of one type of boundary data for every field. For the parity-invariant versions of higher-spin gravity, there exists evidence \cite{Neiman:2014npa} that the answer is yes, at least up to contact terms, to all orders in perturbation theory. Before reviewing this evidence, we will test for the same property at lowest order for conformally-coupled massless interacting scalars, for Yang-Mills theory and for GR. We will find that the property fails to hold in the scalar case, but holds for Yang-Mills and GR at the 3-point level (again, up to contact terms). 

\subsection{Free fields and the role of masslessness} \label{sec:encoding:free}

\subsubsection{Field equations, conformal weights and propagators}

In this subsection, we analyze the interplay between antipodal symmetry and boundary data for free fields. This extends the treatment in \cite{Neiman:2014npa}, which addressed (fully) massless gauge fields. We describe a free bosonic field with spin $s\geq 0$ by a symmetric tensor $\varphi^{\mu_1\dots\mu_s}$, satisfying the equations:
\begin{align}
 \varphi_\nu^{\nu\mu_3\dots\mu_s} = 0 \ ; \quad \nabla_\nu \varphi^{\nu\mu_2\dots\mu_s} = 0 \ ; \quad (\Box - m^2)\varphi^{\mu_1\dots\mu_s} = 0 \ , \label{eq:free_eqs}
\end{align}
where the tracelessness and transversality conditions are only present for spins $s\geq 2$ and $s\geq 1$, respectively. Depending on the mass, they are either genuine field equations or gauge conditions. The d'Alembertian is defined as $\Box = \nabla_\mu\nabla^\mu$. In the spin-0 case, $m^2=0$ corresponds to a minimally-coupled massless scalar, while $m^2=2$ is conformally-coupled massless. For fields with spin $s\geq 1$, the ``mass'' $m$ does not coincide with the intuitive notion of mass from flat space. In particular, the usual notion of ``fully massless'' gauge fields corresponds to $m^2 = 2 + 2s - s^2$. The ``partially massless'' gauge fields of \cite{Deser:2001us}, which are peculiar to de Sitter space, occur at the mass values:
\begin{align}
 m^2 = s + 2 - j(j-1) \ . \label{eq:partially_massless}
\end{align}
Here, the integer $j$ with range $1\leq j \leq s$ is the minimal absolute value in the field's helicity spectrum (the maximal helicity is always the spin $s$). Full masslessness corresponds to $j=s$, so that the only helicities are $\pm s$. Ordinary massive fields have absolute values of the helicity ranging all the way from $0$ to $s$.

The asymptotic boundary data for $\varphi^{\mu_1\dots\mu_s}$ is given by a pair of symmetric traceless fields on e.g. $\scri^+$, with conformal weights (note that these assume raised indices on $\varphi^{\mu_1\dots\mu_s}$):
\begin{align}
 \Delta_\pm = \frac{3}{2} + s \pm \sqrt{s + \frac{9}{4} - m^2} \ . \label{eq:Delta}
\end{align}
For fully massless gauge fields, the boundary data with $\Delta=\Delta_\pm$ correspond respectively to the electric and magnetic parts of the asymptotic field strength.

The conformal weights \eqref{eq:Delta} can be read off from the boundary-to-bulk propagators:
\begin{align}
 \varphi^{\mu_1\dots\mu_s}(x) = \frac{M^{\mu_1\nu_1}x_{\nu_1}\dots M^{\mu_s\nu_s}x_{\nu_s}}{(x\cdot\ell \pm i\varepsilon)^w} \ ; \quad 
 w \in \left\{\Delta_+,\Delta_-\right\} \ . \label{eq:propagator}
\end{align}
Here, $x\in dS_4$ is the bulk point at which the field is evaluated, represented by the unit spacelike vector $x^\mu\in\bbR^{1,4}$. The boundary source point $\ell\in\scri^+$ is represented by a future-pointing null vector $\ell^\mu\in\bbR^{1,4}$. The propagator's polarization is encoded in the totally-null bivector $M^{\mu\nu}$, which has the form:
\begin{align}
 M^{\mu\nu} = 2\ell^{[\mu}\lambda^{\nu]} \ , \quad \text{with $\lambda^\mu$ a complex null vector orthogonal to $\ell^\mu$} \ . \label{eq:M}
\end{align}
The exponent $w$ is the conformal weight of the propagator \eqref{eq:propagator} in the asymptotic limit \eqref{eq:asymptote}, for boundary points $l\neq\ell$. Using \eqref{eq:nabla}, one can verify that the propagator solves the field equations \eqref{eq:free_eqs}, if and only if $w$ coincides with one of the weights \eqref{eq:Delta}. 

Crucially, the various notions of masslessness discussed above coincide with the conformal weights \eqref{eq:Delta} having integer values (since $\Delta_+ + \Delta_- = 3+2s$, either both weights are integers or neither is). For a scalar field, $\Delta_\pm = (3,0)$ corresponds to the minimally-coupled massless case, with $\Delta_\pm = (2,1)$ for conformally-coupled massless. For spin $s\geq 1$, the partially massless case \eqref{eq:partially_massless} corresponds to $\Delta_\pm = (1+s+j,2+s-j)$, with $\Delta_\pm = (1+2s,2)$ for full masslessness. This exhausts all the cases with integer weights, assuming $m^2$ doesn't go negative for $s=0$ or below the fully-massless value for $s\geq 1$.

When the exponent $w$ in the propagator \eqref{eq:propagator} is positive, there is a pole or branch cut at the horizon $x\cdot\ell = 0$. For the conformally-coupled massless scalar and for (partially) massless gauge fields, both weight choices $w = \Delta_\pm$ are positive integers; the propagator then has a pole at the horizon, and the $i\varepsilon$ prescription in \eqref{eq:propagator} is required to define how this pole is bypassed. For massive fields, $w$ is a positive \emph{non}-integer, leading to a branch cut at the horizon; in that case, the $i\varepsilon$ prescription also serves to define the behavior in the half-space $x\cdot\ell < 0$, where we use $(-1\pm i\varepsilon)^w = e^{\pm i\pi w}$. 

We will refer to the propagator \eqref{eq:propagator} with the $i\varepsilon$ prescription as ``Euclidean'', since it is regular on one of the two Euclidean AdS spaces defined by taking $x^\mu$ imaginary:
\begin{align}
 H^\pm &= \left\{ x\in\bbR^{1,4} \,| x_\mu x^\mu = 1\,,\ \Re x^\mu = 0\,,\ \Im x^0 \gtrless 0 \right\} \ . \label{eq:EAdS}
\end{align}

Excluding tachyons, the only case where the exponent $w$ in \eqref{eq:propagator} is \emph{not} positive is the minimally-coupled scalar with the weight choice $w=\Delta_-=0$. The propagator \eqref{eq:propagator} then degenerates into a constant, $\varphi(x) = 1$. In particular, the $\pm i\varepsilon$ prescription becomes redundant, no longer giving two linearly independent variants of the propagator. However, a \emph{different} linearly independent variant exists, given by the sign function:
\begin{align}
 \varphi(x) = \sign(x\cdot\ell) \ . \label{eq:minimal_scalar_sign}
\end{align}
It's easy to check that this propagator also solves the field equation $\Box\varphi = 0$.

\subsubsection{Massive fields (non-integer weights)}

We can superpose propagators of the form \eqref{eq:propagator} into antipodally even/odd combinations:
\begin{align}
 \varphi_\pm^{\mu_1\dots\mu_s}(x) = M^{\mu_1\nu_1}x_{\nu_1}\dots M^{\mu_s\nu_s}x_{\nu_s}
  \left(\frac{1}{(x\cdot\ell + i\varepsilon)^w} \pm \frac{1}{(-x\cdot\ell + i\varepsilon)^w} \right) \ . \label{eq:propagator_pm}
\end{align}
For e.g. $x\cdot\ell > 0$, i.e. to the past of the horizon defined by $\ell$, this evaluates to:
\begin{align}
 \varphi_\pm^{\mu_1\dots\mu_s}(x) = \frac{(1 \pm e^{-i\pi w})}{(x\cdot\ell)^w}\, M^{\mu_1\nu_1}x_{\nu_1}\dots M^{\mu_s\nu_s}x_{\nu_s} \ . \label{eq:propagator_pm_massive}
\end{align}
For massive fields, both weight choices $w = \Delta_\pm$ are non-integers. Then the solutions \eqref{eq:propagator_pm_massive} at $x\cdot\ell > 0$ are nonzero for both choices of $w$ and both antipodal symmetry signs. In particular, for both antipodal symmetry signs, these solutions have nonzero boundary data with conformal weight $w$. Therefore, antipodal symmetry is not associated with the vanishing of either type of boundary data.

\subsubsection{General considerations for massless fields (integer weights)} \label{sec:encoding:free:massless}

For integer $\Delta_\pm$, i.e. massless fields, the previous argument fails, since the antipodally symmetrized propagator \eqref{eq:propagator_pm} now vanishes at $x\cdot\ell \neq 0$ for one of the sign choices. Specifically, the symmetrized propagators become:
\begin{align}
 \varphi_\pm^{\mu_1\dots\mu_s}(x) &\sim \calP\!\left(\frac{1}{(x\cdot\ell)^w}\right) M^{\mu_1\nu_1}x_{\nu_1}\dots M^{\mu_s\nu_s}x_{\nu_s} \ ; \label{eq:PV} \\
 \varphi_\mp^{\mu_1\dots\mu_s}(x) &\sim \delta^{(w-1)}(x\cdot\ell)\, M^{\mu_1\nu_1}x_{\nu_1}\dots M^{\mu_s\nu_s}x_{\nu_s} \ . \label{eq:delta}
\end{align}
Here, $\delta^{(k)}$ is the $k^{\text{th}}$ derivative of the delta function, while the principal value $\cal P$ stands for the average of the two $(x\cdot\ell \pm i\varepsilon)$ prescriptions. Note that the delta-like propagator \eqref{eq:delta} has its support on the horizon $x\cdot\ell = 0$ of the source point $\ell$. For even (odd) $w$, the propagator \eqref{eq:PV} is antipodally even (odd), while \eqref{eq:delta} is antipodally odd (even). 

We will find that antipodal symmetry in the massless case \emph{does} correspond to the vanishing of one type of boundary data, up to a caveat for the minimally-coupled massless scalar. One way to prove this is by directly analyzing the antipodally symmetric propagators \eqref{eq:PV}-\eqref{eq:delta}. Instead, we will present here a more general argument, which uses the explicit propagators only tangentially. A more detailed treatment of the separate kinds of massless fields will be given in sections \ref{sec:encoding:free:conformal_scalar}-\ref{sec:encoding:free:gauge}.

Let us use Poincare coordinates for $dS_4$, which relate to the flat (4+1)d coordinates as:
\begin{align}
 x^\mu = \frac{1}{z}\left(\frac{r^2 - z^2 + 1}{2}, \frac{r^2 - z^2 - 1}{2}, \vec r \right) \ . \label{eq:x}
\end{align}
Here, $z$ is a (past-pointing) conformal time, while $\vec r$ is a flat 3d spatial vector with length $r$. The metric in these coordinates reads:
\begin{align}
 dx_\mu dx^\mu = \frac{-dz^2 + d\vec r\cdot d\vec r}{z^2} \ . \label{eq:Poincare_metric}
\end{align}
Letting $z$ take both positive and negative values, the coordinates \eqref{eq:x} span all of $dS_4$ except the horizon of the boundary point $(1,1,\vec 0)$. The limits $z\rightarrow 0^\pm$ correspond to $\scri^\pm$, respectively. In these limits, the bulk point $x$ asymptotes as in \eqref{eq:asymptote} to the boundary point:
\begin{align}
 l^\mu = \pm\left(\frac{r^2 + 1}{2}, \frac{r^2 - 1}{2}, \vec r \right) \ . \label{eq:l}
\end{align}
The set of null vectors \eqref{eq:l} represents $\scri^\pm$ as a flat section of the $\bbR^{1,4}$ lightcone. The metric on this section is just the flat metric of $\vec r$ space.

We can now make some observations. Since the free field equations are linear, we can consider the evolution of each type of boundary data separately (setting the other type to zero for this purpose). Since the metric is even in $z$, the evolution from $z=0$ of boundary data with weight $\Delta_\pm$ can only generate terms that scale as $z^{\Delta_\pm + 2n}$, with positive integers $n$. Now, we're interested here in the massless case, for which $\Delta_\pm$ are both integers. Furthermore, since they sum to $3+2s$, one of the $\Delta_\pm$ is even, while the other is odd (this is a special property of even bulk dimensions). Thus, one type of boundary data has an even weight and evolves into even powers of $z$, while the other has an odd weight and evolves into odd powers of $z$. 

Our next observation is that in the Poincare coordinates \eqref{eq:x}, the reversal $z\rightarrow -z$ is precisely the antipodal map. Thus, antipodal symmetry is simply parity in $z$. \emph{Assuming} that the solution is regular through $z=0$, this implies that antipodally even (odd) solutions are those with only even (odd) powers of $z$. As we saw above, this in turn is equivalent to the vanishing of boundary data with odd (even) conformal weight. Thus, antipodal symmetry is associated with the vanishing of one type of boundary data (the one with the wrong parity of the conformal weight), \emph{if} one can find a spanning set of solutions that are regular through $z=0$. Geometrically, regularity through $z=0$ is equivalent to regularity upon antipodal identification of the two boundaries $\scri^\pm$.

Thus, our task now is to find suitable spanning sets of solutions. Our starting point is the set of Euclidean boundary-to-bulk propagators \eqref{eq:propagator}, with both weight choices $w=\Delta_\pm$, both $i\varepsilon$ prescriptions, all possible source points $\ell\in\scri^+$ and polarization vectors $\lambda^\mu$. These propagators then span the space of solutions, except for the minimally-coupled massless scalar case, where we must also add the propagator \eqref{eq:minimal_scalar_sign}. For all these propagators, the $z$ dependence near the boundary is governed by the dependence on $x\cdot\ell$. For non-integer weights $w$ in \eqref{eq:propagator}, there is a branch cut at $z=0$, due to the sign flip in $x\cdot\ell$. For integer $w$, i.e. for massless fields, there is no branch cut, except in the minimally-coupled massless scalar propagator \eqref{eq:minimal_scalar_sign}. Instead, the massless propagators have a pole at the horizon $x\cdot\ell = 0$, which intersects the boundary only at the source point $\ell$ and its antipode. Such a pole can be smeared away by integrating over $\ell$, with the $i\varepsilon$ prescription in \eqref{eq:propagator} defining how to bypass the pole in the smearing integral. Equivalently, instead of the Euclidean propagators \eqref{eq:propagator}, one can use their antipodally symmetrized versions \eqref{eq:PV}-\eqref{eq:delta}. 

To reiterate, the propagators \eqref{eq:propagator} or \eqref{eq:PV}-\eqref{eq:delta} can be used as a spanning set of solutions, which become regular through $z=0$ after smearing over $\ell$. Through the argument above, this establishes the relation between antipodal symmetry and the vanishing of boundary data with the ``wrong'' parity of the conformal weight. The exception is the minimally-coupled massless scalar, which requires us to consider also the propagator \eqref{eq:minimal_scalar_sign} with its branch cut at $z=0$. Note that the propagators \eqref{eq:propagator} or \eqref{eq:PV}-\eqref{eq:delta} can be redundant as a spanning set for the solution space. In particular, for gauge fields, the propagators with $w=\Delta_-$ are pure gauge (at least in the fully-massless case, but probably also in the partially-massless one). These and other case-specific details will be discussed below.

\subsubsection{Conformally-coupled massless scalar} \label{sec:encoding:free:conformal_scalar}

Consider first the conformally-coupled massless scalar, i.e. $s=0,m^2=2$. The conformal weights in this case are $\Delta_\pm = 2,1$. By the general argument of section \ref{sec:encoding:free:massless}, antipodally even solutions are those with vanishing $\Delta=1$ boundary data, while antipodally odd solutions are those with vanishing $\Delta=2$ data. The even and odd solutions are spanned most conveniently by the delta-like propagators \eqref{eq:delta} with the two weights $w=\Delta_\pm$:
\begin{align}
 \varphi_+(x) &\sim \delta(x\cdot\ell) \ ; \label{eq:delta_1} \\
 \varphi_-(x) &\sim \delta'(x\cdot\ell) \ . \label{eq:delta_2}
\end{align}
For the antipodally even propagator \eqref{eq:delta_1}, the $\Delta=1$ boundary data vanishes, while the $\Delta=2$ data is given by a delta distribution at the source point $\ell$. Similarly, for the antipodally odd propagator \eqref{eq:delta_2}, the $\Delta=2$ data vanishes, while the $\Delta=1$ data is a delta distribution at $\ell$. These asymptotics are captured by eqs. \eqref{eq:delta_asymptotics}-\eqref{eq:delta_k_asymptotics} in the following theorem, which we prove in Appendix \ref{app:delta_asymptotics}:
\begin{theorem} \label{thm:delta_asymptotics}
 Consider a bulk point $x^\mu\in dS_4$ that asymptotes to a boundary point $l^\mu\in\scri^+$ as in \eqref{eq:asymptote}. Let $\ell^\mu\in\scri^+$ be another boundary point. Then the asymptotics of the step function $\theta(x\cdot\ell)$, the delta function $\delta(x\cdot\ell)$, and its derivatives $\delta^{(k)}(x\cdot\ell)$ with $k\geq 1$, is given by delta distributions on $\scri^+$, as follows:
 \begin{align}
  \int_{\scri^+} d^3\ell\,f(\ell)\,\theta(x\cdot\ell) &= z^3\left(\frac{4\pi}{3}\,f(l) \ + \ \text{positive even powers of }z \right) \ ; \label{eq:theta_asymptotics} \\
  \int_{\scri^+} d^3\ell\,f(\ell)\,\delta(x\cdot\ell) &= z^2\left(4\pi f(l) \ + \ \text{positive even powers of }z \right) \ ; \label{eq:delta_asymptotics} \\
  \begin{split}
    \int_{\scri^+} d^3\ell\,f(\ell)\,\delta^{(k)}(x\cdot\ell) &= z^{2-k}\big(4\pi(-1)^{k-1}(2k-3)!!\,f(l) \\
      &\qquad\qquad\quad + \ \text{positive even powers of }z \big) \ .
  \end{split} \label{eq:delta_k_asymptotics}
 \end{align}
 Here, the integrals are over $\scri^+$, represented by some section of the future-pointing lightcone in $\bbR^{1,4}$. The smearing functions $f(\ell)$ are assumed to be homogeneous of the appropriate degree in $\ell^\mu$, so that the integral is independent of the chosen section. The double factorial $(2k-3)!!$ is defined as $1$ for $k=1$ and $1\cdot 3\cdot 5\cdot\ldots\cdot (2k-3)$ for $k\geq 2$.
\end{theorem}

\subsubsection{Minimally-coupled massless scalar} \label{sec:encoding:free:minimal_scalar}

Consider now the minimally-coupled massless scalar, i.e. $s=0,m^2=0$. The conformal weights in this case are $\Delta_\pm = 3,0$. The antipodally even and odd solutions are spanned most conveniently by the delta-like propagator \eqref{eq:delta} with $w=3$, along with the sign function \eqref{eq:minimal_scalar_sign}:
\begin{align}
 \varphi_+(x) &\sim \delta''(x\cdot\ell) \ ; \label{eq:delta_3} \\
 \varphi_-(x) &\sim \sign(x\cdot\ell) \ . \label{eq:sign_again} 
\end{align}
For the antipodally even propagator \eqref{eq:delta_3}, the $\Delta=3$ boundary data vanishes, while the $\Delta=0$ data is a delta distribution at $\ell$, as can be seen from eq. \eqref{eq:delta_k_asymptotics}. For the antipodally odd propagator \eqref{eq:sign_again}, the $\Delta=0$ data is a constant $\pm 1$, while the $\Delta=3$ data is a delta distribution at $\ell$. To derive this last statement, we separate out the $\Delta=0$ piece on $\scri^+$ through $\sign(x\cdot\ell) = 2\theta(x\cdot\ell) - 1$, and then use eq. \eqref{eq:theta_asymptotics}. 

Note that the propagators \eqref{eq:delta_3}-\eqref{eq:sign_again} indeed span the space of solutions, since we can combine them to produce any profile of both $\Delta=0$ and $\Delta=3$ boundary data. We can therefore read off from these propagators the constraints imposed on the boundary data by antipodal symmetry. From the propagator \eqref{eq:delta_3}, we conclude that antipodally even solutions are the ones with vanishing $\Delta=3$ data. On the other hand, from \eqref{eq:sign_again}, we conclude that antipodally odd solutions are \emph{not} the ones with vanishing $\Delta=0$ data, but rather those for which the $\Delta=0$ data is constrained to $-3/(8\pi)$ times the integral of the $\Delta=3$ data.

Note that by inspecting the principal-value propagator \eqref{eq:PV} with $w=3$, one could have mistakenly deduced that the antipodally odd solutions have vanishing $\Delta=0$ data. That would be incorrect, since this propagator does \emph{not} span the full space of antipodally odd solutions. In fact, one can show that the integral of its $\Delta=3$ data vanishes, so that solutions with this integral non-vanishing are left out.

\subsubsection{Fully and partially massless gauge fields} \label{sec:encoding:free:gauge}

We now turn to (partially) massless gauge fields, with spin $s$ and lowest positive helicity $j$. The conformal weights in this case are $\Delta_- = 2+s-j$ and $\Delta_+ = 1+s+j$. By the general argument of section \ref{sec:encoding:free:massless}, for even (odd) $s-j$, antipodally even (odd) solutions are those with vanishing $\Delta=\Delta_+$ boundary data, while antipodally odd (even) solutions are those with vanishing $\Delta=\Delta_-$ data. 

That being said, it is instructive to discuss precisely how the boundary-to-bulk propagators \eqref{eq:propagator} or \eqref{eq:PV}-\eqref{eq:delta} span the solution space. Let us focus first on the fully massless case $j=s$, where the picture is most clear to us. The first observation is that for $w=\Delta_-$, the Euclidean propagator \eqref{eq:propagator} is pure gauge, and therefore its symmetrized versions \eqref{eq:PV}-\eqref{eq:delta} are pure gauge as well. This can be demonstrated by calculating the gauge-invariant field strength, using e.g. the twistor methods of \cite{Neiman:2014npa}. Thus, the only non-trivial propagators are those with $w=\Delta_+$. One can similarly verify that these are \emph{not} pure gauge (the corresponding field strengths have been calculated e.g. in \cite{Neiman:2014npa}). We conjecture that the same is true in the partially massless case, i.e. that the $w=\Delta_-$ propagators are always pure gauge, while the $w=\Delta_+$ propagators are non-trivial. To attempt a general proof here would take us too far off course. We refer to \cite{Dolan:2001ih,Deser:2003gw,Bekaert:2013zya} for existing treatments of asymptotics and propagators for partially massless fields. 

The above discussion implies that the solution space is spanned by the antipodally symmetrized propagators \eqref{eq:PV}-\eqref{eq:delta} with $w=\Delta_+$:
\begin{align}
 \varphi_\mp^{\mu_1\dots\mu_s}(x) &\sim \calP\!\left(\frac{1}{(x\cdot\ell)^{s+j+1}}\right) M^{\mu_1\nu_1}x_{\nu_1}\dots M^{\mu_s\nu_s}x_{\nu_s} \ ; \label{eq:partially_P} \\
 \varphi_\pm^{\mu_1\dots\mu_s}(x) &\sim \delta^{(s+j)}(x\cdot\ell) M^{\mu_1\nu_1}x_{\nu_1}\dots M^{\mu_s\nu_s}x_{\nu_s} \ . \label{eq:partially_delta}
\end{align}
For even (odd) $s-j$, the delta-like propagator \eqref{eq:partially_delta} is antipodally even (odd), while the principal-value propagator \eqref{eq:partially_P} is antipodally odd (even). As we will see in Theorem \ref{thm:gauge_asymptotics} below, the delta-like propagator \eqref{eq:partially_delta} has vanishing $\Delta=\Delta_+$ data, while its $\Delta=\Delta_-$ data is a delta distribution at the source point $\ell$. The principal-value propagator \eqref{eq:partially_P} has vanishing $\Delta=\Delta_-$ data, while its $\Delta=\Delta_+$ data is some function on $\scri^\pm$. Unlike the scalar case, there is no propagator that yields a simple delta distribution for the $\Delta=\Delta_+$ data. This is to be expected, given the meaning of the two types of boundary data in the gauge-field case. Indeed, for fully massless fields, the $\Delta=\Delta_-$ data describes a magnetic potential on $\scri^\pm$, while the $\Delta=\Delta_+$ data describes an electric field strength. While the magnetic potential is arbitrary, the electric field is constrained by Gauss' law to be divergence-free. Thus, we cannot have a delta distribution in the $\Delta=\Delta_+$ data, since by superposition, that would allow arbitrary electric field profiles. In the partially massless case, the situation is analogous, with Gauss' law replaced by a more complicated constraint with $s-j+1$ derivatives. 

As we will see, for antipodally symmetric solutions of higher-spin gravity, the delta-like propagator \eqref{eq:partially_delta} will suffice. Let us then work out its asymptotics, and see explicitly that it yields a delta distribution in the $\Delta=\Delta_-$ data. Recalling that the polarization bivector $M^{\mu\nu}$ has the structure \eqref{eq:M}, we can write the propagator \eqref{eq:partially_delta} more explicitly as:
\begin{align}
 \varphi_\pm^{\mu_1\dots\mu_s}(x) &\sim G^{\mu_1\dots\mu_n}_{\nu_1\dots\nu_n}(x;\ell)\lambda^{\nu_1}\dots\lambda^{\nu_n} \ ; \label{eq:phi_lambda} \\
 G^{\mu_1\dots\mu_n}_{\nu_1\dots\nu_n}(x;\ell) &= \delta^{(s+j)}(x\cdot\ell) \left(\ell^{\mu_1}x_{\nu_1} - (x\cdot\ell)\delta^{\mu_1}_{\nu_1}\right) \dots
   \big(\ell^{\mu_s}x_{\nu_s} - (x\cdot\ell)\delta^{\mu_s}_{\nu_s}\big) \ , \label{eq:G}
\end{align}
where the ``stripped'' propagator $G^{\mu_1\dots\mu_n}_{\nu_1\dots\nu_n}$ is orthogonal to $x_\mu$ in its lower indices and to $\ell^\nu$ in its upper ones.

Since our propagator's boundary data is distributional, we should smear over the boundary source point $\ell$, treating the polarization factor $\lambda^{\nu_1}\dots\lambda^{\nu_n}$ as a smearing function $\lambda^{\nu_1}(\ell)\dots\lambda^{\nu_n}(\ell)$. The asymptotics of the propagator \eqref{eq:phi_lambda}-\eqref{eq:G} is then given by the following formula, which we prove in Appendix \ref{app:delta_asymptotics}.

\begin{theorem} \label{thm:gauge_asymptotics}
 Consider a bulk point $x^\mu\in dS_4$ that asymptotes to a boundary point $l^\mu\in\scri^+$ as in \eqref{eq:asymptote}. Let $\ell^\mu\in\scri^+$ be another boundary point. Then the asymptotics of the (partially) massless gauge field propagator \eqref{eq:G} is given by a delta distribution on $\scri^+$, as follows:
 \begin{align}
   \begin{split}
     \int_{\scri^+} d^3\ell\,&\lambda^{\nu_1}(\ell)\dots\lambda^{\nu_s}(\ell)\, G^{\mu_1\dots\mu_n}_{\nu_1\dots\nu_n}(x;\ell) 
      = z^{2-j}\big(C(s,j)\,\lambda^{\mu_1}(l)\dots\lambda^{\mu_s}(l) \\
      &+ \ \text{positive even powers of }z \big)
      + \ l^\mu\text{-parallel terms of the form } l^{(\mu_1}f^{\mu_2\dots\mu_s)} \ , 
   \end{split} \label{eq:G_asymptotics} 
 \end{align}
 where $C(s,j)$ is a numerical coefficient given by:
 \begin{align}
  C(s,j) = 4\pi(-1)^{j+1}s!(s+j)!(2j-3)!! \sum_{k=0}^s \frac{(-1)^k}{(s-k)!(j+k)!} \ . \label{eq:C}
 \end{align}
 The integral in \eqref{eq:G_asymptotics} is over $\scri^+$, represented by some section of the future-pointing lightcone in $\bbR^{1,4}$. The complex polarization vectors $\lambda^\mu(\ell)$ are taken to be null and orthogonal to $\ell^\mu$. We assume that they're homogeneous functions of $\ell^\mu$ of degree $(j-2)/s$, so that the integral is   independent of the chosen section.
\end{theorem}
For the fully massless case $j=s$, the sum in \eqref{eq:C} can be evaluated as $1/\left(2(s!)^2\right)$, using the binomial expansion of $0=(1-1)^{2s}$. The numerical coefficient then simplifies to:
\begin{align}
 C(s,s) = 2\pi(-1)^{s+1}\frac{(2s)!(2s-3)!!}{s!} \ .
\end{align}

\subsection{Interacting massless theories} \label{sec:encoding:interacting}

Among free field theories, we've seen that for massless fields, antipodal symmetry has a simple encoding on the boundary. Apart from a subtlety in the minimally-coupled scalar case, this encoding is just the vanishing of one type of boundary data. Let us now test for the analogous property in \emph{interacting} massless theories. Our primary interest is the special case of higher-spin gravity, which we take up in section \ref{sec:encoding:interacting:HS}.

\subsubsection{Generalities} \label{sec:encoding:interacting:general}

We recall that only parity-invariant interactions allow for antipodally symmetric solutions. Thus, we can restrict attention to theories where the fields transform as either tensors or pseudotensors under parity. In fact, we will further restrict to theories where the spin-0 fields are either scalars or pseudoscalars, while the gauge fields with spin $s\geq 1$ are all tensors. We take the spin-0 fields to be conformally-coupled massless (rather than minimally-coupled), and the gauge fields to be fully massless (rather than partially). With these restrictions, our discussion will still encompass the parity-invariant versions of higher-spin theory, as well as Yang-Mills theory and GR. 

Since a field's parity determines its antipodal symmetry sign (if any), we are led to studying solutions where the gauge fields are antipodally even, while the scalars/pseudoscalars are antipodally even/odd, respectively. As it happens, these are precisely the cases that admit delta-like boundary-to-bulk propagators with support on the source point's horizon. From eqs. \eqref{eq:delta_1}-\eqref{eq:delta_2} and \eqref{eq:partially_delta}, substituting $j=s$ in the latter, these propagators read:
\begin{align}
 \varphi_+^{\mu_1\dots\mu_s}(x) &\sim \delta^{(2s)}(x\cdot\ell) M^{\mu_1\nu_1}x_{\nu_1}\dots M^{\mu_s\nu_s}x_{\nu_s} \ ; \label{eq:even_propagator} \\
 \varphi_-(x) &\sim \delta'(x\cdot\ell) \ . \label{eq:odd_propagator}
\end{align}
Here, \eqref{eq:even_propagator} is the antipodally even propagator for all spins (including the scalar), while \eqref{eq:odd_propagator} is the antipodally odd propagator for the pseudoscalar.

At the free-field level, the solution space for antipodally even fields is spanned by the propagator \eqref{eq:even_propagator}. As we've seen, these solutions have only the $\Delta=2$ boundary data non-vanishing in the scalar case, and only the magnetic data non-vanishing for gauge fields with spin $s\geq 1$. Similarly, the antipodally odd pseudoscalar solutions are spanned by the propagator \eqref{eq:odd_propagator}, and have only the boundary data with $\Delta=1$ non-vanishing. For brevity, we will refer to the types of boundary data that are non-vanishing in the free theory as ``normal'', while the vanishing ones will be called ``abnormal''. 

In general, the ``abnormal'' boundary data for a solution in $dS_4/\bbZ_2$ can be identified with the derivative of the on-shell action $S$ with respect to the ``normal'' data. In turn, the on-shell action as a functional of ``normal'' data can be expanded in a Taylor series, with the $n^{\text{th}}$-order coefficient given by an asymptotic $n$-point function with ``normal'' data as boundary conditions. In this language, the vanishing of the ``abnormal'' data in free massless theory can be rephrased as the vanishing of the 2-point function in $dS_4/\bbZ_2$ with ``normal'' data as boundary conditions. Similarly, the vanishing of the ``abnormal'' data in \emph{interacting} theory is perturbatively equivalent to the vanishing of all higher $n$-point functions. It is this property that we will test for in this section, for various interacting massless theories. For simplicity, we will focus on $n$-point functions with $n$ distinct boundary points, i.e. we will neglect contact terms.

The classical $n$-point functions in question can be found by evaluating tree-level Witten diagrams in $dS_4/\bbZ_2$. We will focus on single-vertex diagrams, consisting of $n$ boundary-to-bulk propagators and a single interaction vertex in the bulk. In higher-spin theory, as argued in \cite{Didenko:2012tv}, such diagrams should be sufficient for evaluating all $n$-point functions at separated points (which will allow us to show that the latter vanish in the $dS_4/\bbZ_2$ setup). In conventional field theories, single-vertex diagrams are sufficient only for the first non-trivial $n$-point functions. For scalar interactions, we will see that already these lowest $n$-point functions are non-vanishing, i.e. that interactions spoil the simple relation between boundary data and antipodal symmetry already at lowest order. For Yang-Mills and GR, we will find that the lowest $n$-point functions (with $n=3$) vanish, though we suspect that higher $n$-point functions will not.

Our single-vertex $n$-point diagrams in $dS_4/\bbZ_2$ can be evaluated by a calculation in $dS_4$, with the antipodal symmetry encoded in the choice of propagators \eqref{eq:even_propagator}-\eqref{eq:odd_propagator}. We recall that these propagators can be expressed as the \emph{difference} between two ``Euclidean'' propagators \eqref{eq:propagator} with opposite $\pm i\varepsilon$ prescriptions. This is in contrast to standard (A)dS/CFT calculations (see e.g. \cite{Freedman:1998tz,Arutyunov:1999nw,Maldacena:2002vr}), where one uses the ``Euclidean'' propagator itself. Thus, the $n$-point functions that we're after are distinct from the ones that are usually calculated; in fact, they are simpler, due to the delta-like form of the propagators \eqref{eq:even_propagator}-\eqref{eq:odd_propagator}.

\subsubsection{Interacting scalars and pseudoscalars}

Consider an interacting conformally-coupled massless spin-0 field, governed by the Lagrangian:
\begin{align}
 \mathcal{L} = -\frac{1}{2} \nabla_\mu \varphi \nabla^\mu \varphi - \varphi^2 - \frac{\lambda_3}{3!} \varphi^3 - \frac{\lambda_4}{4!} \varphi^4 \ .
\end{align}
In the presence of a cubic coupling $\lambda_3\neq 0$, the field $\varphi$ must be a scalar, i.e. it must have even parity. The theory then admits antipodally even solutions in $dS_4$. On the other hand, if $\lambda_3$ vanishes, then $\varphi$ can be either scalar or pseudoscalar, allowing for antipodal symmetry of either sign.

As discussed in section \ref{sec:encoding:interacting:general}, we wish to know whether the tree-level $n$-point functions for these theories in $dS_4/\bbZ_2$ vanish. We will show that for scalars with $\lambda_3 \neq 0$, the 3-point function is nonzero, while for both scalars and pseudoscalars with $\lambda_3 = 0$ and $\lambda_4 \neq 0$, the 4-point function is nonzero. This implies that the interactions spoil the simple relationship between antipodal symmetry and boundary data, already at leading order. 

Let us first consider a scalar $\varphi$ with $\lambda_3\neq 0$. We wish to calculate the 3-point function, which is given by a single-vertex diagram with three boundary-to-bulk propagators. By $SO(1,4)$ symmetry, any triple of distinct source points on $\scri$ is equivalent to any other. For concreteness, let us choose three points parameterized by the following null vectors in the $\bbR^{1,4}$ embedding space: 
\begin{align}
 \ell_1^\mu = (1,1,0,0,0) \ ; \quad \ell_2^\mu = (1,-1,0,0,0) \ ; \quad \ell_3^\mu = (1,0,1,0,0) \ . \label{eq:source_123}
\end{align}
We represent the bulk interaction point by a unit spacelike vector $x^\mu\in\bbR^{1,4}$. This will be integrated over using the 4-volume measure on $dS_4$, which can be written as: 
\begin{align}
 dV_4 = 2\delta(x\cdot x - 1)d^5x \ . \label{eq:dS_volume}
\end{align}
In $dS_4/\bbZ_2$, the integration range is effectively halved, which can be taken into account by dropping the factor of $2$ in \eqref{eq:dS_volume}.

It remains to plug in the even boundary-to-bulk propagator \eqref{eq:delta_1}. Up to an overall numerical factor, the 3-point function then reads:
\begin{align}
 S^{(3)}(\ell_1;\ell_2;\ell_3) \sim \lambda_3 \int d^5 x\, \delta (x \cdot \ell_1) \delta (x \cdot \ell_2)\delta (x \cdot \ell_3) \delta(x\cdot x -1) \ .
\end{align}
With the choice of source points \eqref{eq:source_123}, this evaluates to:
\begin{align}
 \begin{split}
   S^{(3)}(\ell_1;\ell_2;\ell_3) \sim \lambda_3\int d^5 x \, &\delta (-x^0 + x^1)\,\delta (-x^0-x^1)\,\delta (-x^0+x^2) \\
      &\times \delta\left(-(x^0)^2+(x^1)^2+(x^2)^2+(x^3)^2+(x^4)^2 -1\right) = \frac{\pi\lambda_3}{2} \ .
 \end{split}
\end{align}
Thus, the 3-point function is nonzero.

Let us turn now to a scalar $\varphi$ with $\lambda_3 = 0$. The 3-point function then vanishes, while the 4-point function is given by a single-vertex diagram with four boundary-to-bulk propagators. Unlike in the 3-point case, not all choices of four source points are equivalent under $SO(1,4)$. However, it suffices to find \emph{one} choice of points for which the 4-point function is nonzero. Consider, then, the three points from \eqref{eq:source_123}, together with:
\begin{align}
 \ell_4^\mu = (1,0,0,1,0) \ . \label{eq:source_4}
\end{align}
Up to an overall numerical factor, the 4-point function now reads:
\begin{align}
 \begin{split}
   S^{(4)}(\ell_1;\ell_2;\ell_3;\ell_4) \sim \lambda_4 \int &d^5 x\, \delta(-x^0 + x^1)\,\delta(-x^0-x^1)\, \delta(-x^0+x^2)\, \delta(-x^0+x^3) \\  
     &\times \delta(-(x^0)^2+(x^1)^2+(x^2)^2+(x^3)^2+(x^4)^2 -1) = \frac{\lambda_4}{2} \ .
 \end{split} 
\end{align}
The result is again nonzero.

Finally, consider a pseudoscalar $\varphi$, which necessarily has $\lambda_3 = 0$. We should now use the odd boundary-to-bulk propagator \eqref{eq:delta_2}. With the same choice of source points \eqref{eq:source_123},\eqref{eq:source_4} as above, the 4-point function reads, up to an overall numerical factor:
\begin{align}
 \begin{split}
   S^{(4)}(\ell_1;\ell_2;\ell_3;\ell_4) \sim \lambda_4 \int &d^5 x\, \delta' (-x^0 + x^1)\,\delta' (-x^0-x^1)\, \delta' (-x^0+x^2)\, \delta'(-x^0+x^3) \\  
     &\times \delta(-(x^0)^2+(x^1)^2+(x^2)^2+(x^3)^2+(x^4)^2 -1) = -\frac{3\lambda_4}{16} \ .
 \end{split}
\end{align}
Thus, the 4-point function is again nonzero.

\subsubsection{Yang-Mills theory} \label{sec:encoding:interacting:YM}

We now turn to Yang-Mills theory, described by the Lagrangian:
\begin{equation}
 \mathcal{L} = -\frac{1}{4} F^a_{\mu \nu} F_a^{\mu \nu} \ ; \quad F^a_{\mu\nu} = 2\del_{[\mu}A_{\nu]}^a + f^{a}{}_{bc}A_\mu^b A_\nu^c \ . \label{eq:L_YM}
\end{equation}
Here, $(a,b,c,\dots)$ are indices in the gauge algebra, with structure constants $f_{abc}$. Due to the cubic interaction in \eqref{eq:L_YM}, $A_\mu^a$ must be a vector rather than a pseudovector. Therefore, the theory admits antipodally even solutions in $dS_4$. 

Our goal is to calculate the tree-level 3-point function for the theory on $dS_4/\bbZ_2$. This 3-point function is encoded in a diagram with three boundary-to-bulk propagators and an interaction vertex arising from the cubic piece of \eqref{eq:L_YM}:
\begin{align}
 \mathcal{L}^{(3)} \sim f_{abc}\,\nabla_\mu A_\nu^a\, A^{\mu b} A^{\nu c} \ . \label{eq:L_3_YM}
\end{align}
We will evaluate this diagram for separated source points, and find that it vanishes. Thus, the relationship between antipodal symmetry and boundary data is preserved to leading order in the interaction, at least up to contact terms.

To evaluate the diagram, we choose the three boundary source points $\ell_i$ from \eqref{eq:source_123}, recalling that all choices of non-coincident points are equivalent under $SO(1,4)$. In addition, for each $i=1,2,3$, we choose a polarization vector $\lambda_i^\mu$ orthogonal to $\ell_i^\mu$. We suppress the sources' color factors, since they simply yield a constant when contracted with $f_{abc}$. 

The even propagator for spin-1 gauge fields can be obtained by setting $s=1$ in \eqref{eq:phi_lambda}-\eqref{eq:G}:
\begin{align}
 A^\mu(x;\ell,\lambda) \sim \delta''(x \cdot \ell) \left((x \cdot \lambda) \ell^\mu - (x \cdot \ell) \lambda^\mu \right) \ . \label{eq:A}
\end{align}
From this we can derive the curl $\nabla_{[\mu}A_{\nu]}$, which appears in \eqref{eq:L_3_YM}:
\begin{align}
 \nabla_{[\mu}A_{\nu]}(x;\ell,\lambda) \sim \delta''(x\cdot\ell) P_\mu^\rho(x) P_\nu^\sigma(x) \ell^{[\rho}\lambda^{\sigma]} \ . \label{eq:curl_A}
\end{align}
Here, $P_\mu^\nu(x)$ is the projector \eqref{eq:projector} into the $dS_4$ tangent space at $x$. The projectors in \eqref{eq:curl_A} can in fact be discarded, since the indices of $\nabla_{[\mu}A_{\nu]}$ in \eqref{eq:L_3_YM} are contracted with factors of $A^\mu$, which already lie in the tangent space. The only remaining $x$ dependence in \eqref{eq:curl_A} is in the factor $\delta''(x\cdot\ell)$. Thus, the overall $x$ dependence in the vertex \eqref{eq:L_3_YM} consists of a $\delta''(x\cdot\ell)$ factor for each leg, along with a polarization factor linear in $x$, as in \eqref{eq:A}, for two of the legs.

Putting everything together, the 3-point function takes the form:
\begin{align}
 \begin{split}
   S^{(3)}(\ell_1,\lambda_1;\ell_2,\lambda_2;\ell_3,\lambda_3) \sim \int d^5x\, &C(x;\lambda_1,\lambda_2,\lambda_3)\, \delta(x\cdot x - 1) \\ 
    &\times \delta''(-x^0+x^1)\, \delta''(-x^0-x^1)\,\delta''(-x^0+x^2) \ ,
 \end{split} \label{eq:S_3_YM}
\end{align}
where $C$ is a polarization factor quadratic in $x^\mu$. Terms in $C(x)$ that are odd in either $x^3$ or $x^4$ will integrate to zero, so we can replace $C(x)$ with:
\begin{align}
 C(x) \, \longrightarrow \, a (x^3)^2 + b (x^4)^2 + c(x^0,x^1,x^2) \ ,
\end{align}
where $a$ and $b$ are constants, while $c$ is a quadratic function. One can now explicitly evaluate the integral \eqref{eq:S_3_YM}, and find that it vanishes. 

We conclude that the vanishing of the ``abnormal'' boundary data in $dS_4/\bbZ_2$ persists to second order in the ``normal'' data, or, equivalently, to first order in the Yang-Mills coupling, at least up to contact terms.

\subsubsection{General Relativity} \label{sec:encoding:interacting:GR}

We now turn to General Relativity, treated perturbatively over a $dS_4$ or $dS_4/\bbZ_2$ background. With our normalization $\Lambda=3$ of the cosmological constant, the Einstein-Hilbert action reads:
\begin{align}
 S = \frac{1}{2}\int d^4x\, \sqrt{-g}\,(R-6) \ , \label{eq:S_GR}
\end{align}
where $g$ is the determinant of the metric $g_{\mu\nu}$, and $R$ is the Ricci scalar. The metric in the action \eqref{eq:S_GR} must be a tensor rather than a pseudotensor. Therefore, we will be interested in antipodally even solutions in $dS_4$. As in the Yang-Mills case, our goal will be to calculate the tree-level 3-point function in $dS_4/\bbZ_2$ at non-coincident points. We will again find that this 3-point function vanishes. 

As a first step, we expand the metric as $g_{\mu\nu} = g^{(0)}_{\mu\nu} + h_{\mu\nu}$, where $g^{(0)}_{\mu\nu}$ is the pure $dS_4$ or $dS_4/\bbZ_2$ metric $g^{(0)}_{\mu\nu} = P_{\mu\nu}(x)$, while $h_{\mu\nu}$ is the dynamical perturbation. From now on, covariant derivatives and the raising/lowering of indices will be done with respect to the background metric.

Our 3-point function is again found by evaluating a single-vertex diagram with three boundary-to-bulk propagators. These propagators can be found by setting $s=2$ in \eqref{eq:phi_lambda}-\eqref{eq:G}:
\begin{align}
 h^{\mu\nu}(x;\ell,\lambda) \sim \delta''''(x \cdot \ell) \left((x \cdot \lambda) \ell^\mu - (x \cdot \ell) \lambda^\mu \right)
 \left((x \cdot \lambda) \ell^\nu - (x \cdot \ell) \lambda^\nu \right) \ . \label{eq:h}
\end{align}
Here, $\ell$ is a boundary source point, while $\lambda^\mu$ is a null polarization vector orthogonal to $\ell^\mu$. We should now plug three propagators of the form \eqref{eq:h} into the cubic piece of the Einstein-Hilbert action \eqref{eq:S_GR}. Note that boundary terms can be discarded, since for non-coincident boundary points, the three propagators are never non-vanishing at the same point on $\scri$. This is in contrast to the situation with the Euclidean propagators in \cite{Arutyunov:1999nw}. In addition, we can simplify the action's cubic piece by using the linearized gauge conditions and field equations \eqref{eq:free_eqs} satisfied by our boundary-to-bulk propagators:
\begin{align}
 h_\mu^\mu = 0 \ ; \quad \nabla_\nu h^\nu_\mu = 0 \ ; \quad \Box h_{\mu\nu} = 2h_{\mu\nu} \ . \label{eq:free_h}
\end{align}
After these simplifications, the cubic piece of the action becomes an integral over $dS_4/\bbZ_2$ of the following Lagrangian:
\begin{align}
 \mathcal{L}^{(3)} = \frac{3}{4} h^\mu_\nu h^\nu_\rho h^\rho_\mu + \frac{3}{8} h^{\mu \nu}\nabla_\mu h_{\rho\sigma}\nabla_\nu h^{\rho\sigma} \ . \label{eq:L_3_GR}
\end{align}
Let us now plug into this vertex three propagators of the form \eqref{eq:h}, with the boundary source points $\ell_1,\ell_2,\ell_3$ from \eqref{eq:source_123} and arbitrary polarizations $\lambda_1^\mu,\lambda_2^\mu,\lambda_3^\mu$. Focusing first on the $h^\mu_\nu h^\nu_\rho h^\rho_\mu$ piece of the vertex \eqref{eq:L_3_GR}, we get a contribution to the 3-point function proportional to:
\begin{align}
 \int d^5x\, &C(x;\lambda_1,\lambda_2,\lambda_3)\, \delta(x\cdot x - 1)\, \delta''''(-x^0+x^1)\, \delta''''(-x^0-x^1)\, \delta''''(-x^0+x^2) \ , \label{eq:S_3_GR}
\end{align}
where the polarization factor $C$ is a polynomial in $x^\mu$ of order no higher than $6$. We can get rid of the first delta function in \eqref{eq:S_3_GR} by performing the integrals over $x^3,x^4$, which leave the $\delta''''$ factors untouched. Denoting $(x^0,x^1,x^2)\equiv\mathbf{x}$ and $(x^3,x^4)\equiv(\rho\cos\phi,\rho\sin\phi)$, these integrals affect the $C$ and $\delta(x\cdot x - 1)$ factors as follows:
\begin{align}
 \begin{split}
  \int dx^3 dx^4 \,C(x)&\,\delta(x \cdot x -1) = \int_0^\infty \rho\, d\rho \int_0^{2\pi} d\phi\,C(\mathbf{x},\rho\cos\phi,\rho\cos\phi)\, 
      \delta(\mathbf{x}\cdot\mathbf{x}+\rho^2-1) \\
   &= \frac{1}{2}\theta(1-\mathbf{x}\cdot\mathbf{x})\int_0^{2\pi} d\phi\, 
       C\big(\mathbf{x},\sqrt{1-\mathbf{x}\cdot\mathbf{x}}\cdot\cos\phi,\sqrt{1-\mathbf{x}\cdot\mathbf{x}}\cdot\sin\phi\big) \ ,
 \end{split}
\end{align}
where $\theta$ is the step function. Only terms with even powers of $\sqrt{1-\mathbf{x}\cdot\mathbf{x}}$ will survive the $d\phi$ integral. These terms can be reorganized into a polynomial $Q(\mathbf{x})$ of order no higher than that of $C(x)$, i.e. at most $6$:
\begin{align}
 \int dx^3 dx^4 \,C(x)&\,\delta(x \cdot x -1) = \theta(1-\mathbf{x}\cdot\mathbf{x})\,Q(\mathbf{x}) \ .
\end{align}
Let us now plug this into the 3-point function contribution \eqref{eq:S_3_GR}, which becomes:
\begin{align}
 \int d^3\mathbf{x}\, &Q(\mathbf{x};\lambda_1,\lambda_2,\lambda_3)\, \theta(1-\mathbf{x}\cdot\mathbf{x})\, \delta''''(-x^0+x^1)\, \delta''''(-x^0-x^1)\, \delta''''(-x^0+x^2) \ . \label{eq:S_3_GR_partially_integrated}
\end{align}
The three $\delta''''$ factors have mutual support only at the origin of $\mathbf{x}$ space. Therefore, we can replace $\theta(1-\mathbf{x}\cdot\mathbf{x})\rightarrow 1$. We are then left with an integral of three $\delta''''$ factors against a polynomial of order no higher than $6$. Since the number of derivatives on the delta functions exceeds the order of the polynomial, this integral vanishes. Thus, the $h^\mu_\nu h^\nu_\rho h^\rho_\mu$ piece of the vertex \eqref{eq:L_3_GR} does not contribute to the 3-point function.

The $h^{\mu \nu}\nabla_\mu h_{\rho\sigma}\nabla_\nu h^{\rho\sigma}$ piece of the vertex \eqref{eq:L_3_GR} can be treated similarly. First, we use \eqref{eq:nabla} to express the covariant derivatives in terms of flat (4+1)d derivatives:
\begin{align}
 h^{\mu \nu}\nabla_\mu h_{\rho\sigma}\nabla_\nu h^{\rho\sigma} = h^{\mu \nu} P^{\kappa\rho}(x) P^{\lambda\sigma}(x) \del_\mu h_{\kappa\lambda}\del_\nu h_{\rho\sigma} \ ,
 \label{eq:deriv_vertex}
\end{align}
where $P^{\kappa\rho}(x)$ is the projector \eqref{eq:projector} into the $dS_4$ tangent space. Note that the $\mu\nu$ indices in \eqref{eq:deriv_vertex} do not require projectors, since $h^{\mu\nu}$ is already inside the tangent space. 

The calculation now proceeds analogously to the one for the $h^\mu_\nu h^\nu_\rho h^\rho_\mu$ vertex. After performing the $dx^3 dx^4$ integrals, we end up with expressions like \eqref{eq:S_3_GR_partially_integrated}, but with different numbers of derivatives on the delta functions and a different order for the polynomial $Q(\mathbf{x})$. Specifically, each projector in \eqref{eq:deriv_vertex} raises the order of the polynomial by at most $2$, while each flat derivative either adds a derivative onto one of the delta functions or reduces the order of the polynomial by $1$. It's easy to see that the number of derivatives on the delta functions still exceeds the order of the polynomial, and therefore this contribution to the 3-point function also vanishes.

We conclude that the 3-point function at non-coincident points for GR on a $dS_4/\bbZ_2$ background is zero. Thus, the vanishing of the ``abnormal'' boundary data for antipodally symmetric solutions persists to second order in the ``normal'' data, or, equivalently, to first order in Newton's constant, at least up to contact terms.

\subsubsection{Higher-spin gravity} \label{sec:encoding:interacting:HS}

Let us now turn to the true case of interest for dS/CFT, i.e. higher-spin theory. The linearized limit of higher-spin theory consists of a conformally-coupled massless spin-0 field, together with an infinite tower of fully massless gauge fields with increasing spin. For physics in $dS_4/\bbZ_2$, we are only interested in parity-invariant versions of the theory. There are two such versions, known as the type-A and type-B models. The two models assign respectively even and odd parity to the zero-form master field $B$ that contains the field strengths of all spins. When decomposed into component fields, this implies that the spin-0 field is respectively a scalar or a pseudoscalar, while the gauge fields with spin $s\geq 1$ are always tensors. Therefore, the type-A (type-B) model allows for antipodally symmetric solutions in $dS_4$ with an antipodally even (odd) spin-0 field and antipodally even gauge fields. These are precisely the cases covered by the delta-like boundary-to-bulk propagators \eqref{eq:even_propagator}-\eqref{eq:odd_propagator}. 

Once again, our goal is to see whether the vanishing of the ``abnormal'' boundary data for antipodally symmetric solutions persists beyond the free-field level. Partial evidence from \cite{Neiman:2014npa} suggests that the answer is yes, to all orders in perturbation theory. We now turn to review this evidence. We recall that the vanishing of the ``abnormal'' data is equivalent to the vanishing of (tree-level) $n$-point functions in $dS_4/\bbZ_2$ with boundary conditions that fix the ``normal'' data. Note that such boundary conditions in the type-A and type-B higher-spin models are precisely the ones that preserve higher-spin symmetry \cite{Vasiliev:2012vf}, and that lead (in standard AdS holography) to a free CFT as the holographic dual. 

To evaluate the $n$-point functions in $dS_4/\bbZ_2$, we plug into the existing bulk methods \cite{Giombi:2009wh,Giombi:2010vg,Colombo:2012jx,Didenko:2012tv,Didenko:2013bj} for calculating $n$-point functions in (Euclidean) AdS. The transition from Euclidean AdS to $dS_4/\bbZ_2$ can be expressed as a modification of the boundary-to-bulk propagators. As a first step, following \cite{Anninos:2011ui}, we analytically continue from Euclidean $AdS_4$ to Lorentzian $dS_4$, represented as in \eqref{eq:EAdS} and \eqref{eq:dS}, respectively. This leads to the ``Euclidean'' propagators \eqref{eq:propagator} in $dS_4$. We then take the difference between the two $\pm i\varepsilon$ prescriptions in \eqref{eq:propagator}, resulting in the antipodally symmetric propagators \eqref{eq:even_propagator}-\eqref{eq:odd_propagator}. After modifying the propagators in this way, one can show that the $n$-point functions indeed vanish, to the extent that they can be calculated using existing bulk methods. In particular, so far these methods can extract only the $n$-point functions for non-coincident points.

The case for which a full bulk calculation exists is the 3-point function with non-coincident points \cite{Giombi:2009wh,Giombi:2010vg}. The more efficient version of this calculation \cite{Giombi:2010vg} uses the fact that on-shell, the higher-spin gauge connection is flat, while the covariant derivatives of the other master fields vanish. One can therefore calculate the master fields at any one bulk point, and then propagate the result to any other by a gauge transformation. The method of \cite{Giombi:2010vg} is then to introduce sources at two boundary points, calculate the master fields to second order at an arbitrary bulk point, and propagate from there to the third boundary point. Now, with the antipodally symmetric propagators \eqref{eq:even_propagator}-\eqref{eq:odd_propagator}, this procedure trivially gives zero, as long as the chosen bulk point is away from the source points' horizons, i.e. away from the support of the delta functions in \eqref{eq:even_propagator}-\eqref{eq:odd_propagator}. Put differently, one can carry out the calculation of \cite{Giombi:2010vg} with a Euclidean propagator with either sign of the $\pm i\varepsilon$ prescription, and get identical results away from the horizons. The vanishing result in $dS_4/\bbZ_2$ then comes from taking the difference between the two $\pm i\varepsilon$ prescriptions.

For $n$-point functions with $n>3$, there is as yet no explicit calculation using the non-linear higher-spin field equations. However, there is an alternative method \cite{Colombo:2012jx}, brought to fruition in \cite{Didenko:2012tv,Didenko:2013bj}, which obtains the result for non-coincident boundary points through a calculation in the language of the linearized bulk theory. In this method, the $n$-point function is obtained as a gauge-invariant scalar product of $n$ boundary-to-bulk propagators at a bulk point. The bulk point is again arbitrary, since the propagators' values at different points are related by gauge transformations. Upon replacing the Euclidean propagators with the antipodally symmetric ones, this procedure again immediately yields zero, as long as the bulk point is chosen away from the horizons of all $n$ source points.

To conclude, there is substantial evidence that the vanishing of the ``abnormal'' type of boundary data in an antipodally symmetric solution persists order by order in perturbation theory, at least up to contact terms. This implies that the simple boundary encoding of a solution's antipodal symmetry persists to all orders in the interaction. The mechanism behind this all-orders result appears unique to higher-spin theory. 

\section{Quantum states for fields in $dS_4/\bbZ_2$} \label{sec:QFT}

In this section, we address the basics of \emph{quantum} field theory in $dS_4/\bbZ_2$. What makes the topic non-trivial is that $dS_4/\bbZ_2$ is globally non-orientable in time. It was pointed out in \cite{Parikh:2002py,Hackl:2014txa} that this precludes a global, observer-independent Hilbert space. Instead, we have a separate quantum description for each observer, based on a partial, observer-dependent notion of time orientation. The challenge then is to understand how the world-pictures of different observers relate to one another. This issue was taken up previously by one of the authors in \cite{Hackl:2014txa}. There, the discussion hinged on the classical phase space, which was shown to have no dS-invariant symplectic structure, but admitted an observer-dependent one. After quantizing with this observer-dependent structure, one could use the Wigner-Weyl transform \cite{Wigner} to map between functionals on the shared phase space and operators on the observer-dependent Hilbert space. This resulted in a recipe for relating the quantum descriptions of different observers, with an underlying global structure provided by phase space functionals. A key feature of this recipe is that pure states do not map into pure states, in contradiction to the claim in \cite{Parikh:2002py}.

Here, we will clarify and mostly vindicate the proposal of \cite{Hackl:2014txa}. We will show that the recipe of \cite{Hackl:2014txa} for translating between observers has a clear physical meaning. We will also show that two observers under this recipe will agree on expectation values of observables in regions that 1) lie in both their causal patches, and 2) are assigned the same time orientation by both observers. The key to our results is to identify the global structure underlying the observers' world-pictures as (a real slice of) the Hilbert space on ordinary $dS_4$. Thus, while $dS_4/\bbZ_2$ is to be thought of as the physical spacetime, we re-introduce its double cover $dS_4$ as a sort of thermofield double \cite{Goheer:2002vf}.

Where we depart from the proposal of \cite{Hackl:2014txa} is in the choice of hypersurface on which the phase space is parameterized. While the recipe of \cite{Hackl:2014txa} uses $\scri$, we will find that the construction has a clear physical meaning if we instead use a Cauchy slice in the causal patch, or, as a limiting case, one of the horizons. The link to data on $\scri$ as proposed in \cite{Hackl:2014txa} is fully clear to us only in the case of free massless fields; this includes the conformally-coupled massless scalar, which was used in \cite{Hackl:2014txa} as an example. We take up the free massless case in section \ref{sec:dS/CFT}. 

For simplicity, we will consider a theory with a single, real spin-0 field $\varphi$ of intrinsic parity $\eta=\pm 1$. It should be clear, however, that the discussion is applicable to more general parity-invariant theories.

\subsection{From states in $dS_4$ to operators in the causal patch} \label{sec:QFT:dS}

As motivated above, we first consider the field theory on $dS_4$, which we view as a double cover of $dS_4/\bbZ_2$. In $dS_4$, there are no problems with time orientation, and the theory can be quantized as usual. In this quantum theory, there exists an antiunitary symmetry operator $\calA$ of the CT type which realizes the antipodal map. As we will see, this operator can be used to map between state vectors in the global $dS_4$ Hilbert space and operators in the Hilbert space of an observer's causal patch. In the application to $dS_4/\bbZ_2$, the latter will be reinterpreted as the Hilbert space of the entire spacetime \emph{as viewed by the observer}. 

In addition, it will be useful for the $dS_4/\bbZ_2$ application to express the global $dS_4$ states in an antipodally symmetric basis. Thus, in this subsection, we construct a map between global states in an antipodally symmetric basis and operators in the causal patch. In section \ref{sec:dS/CFT}, we will eventually use this map to encode the observer's quantum theory in terms of holographic quantities on $\scri$. 

\subsubsection{General framework}

Recall that an observer in $dS_4$ defines a causal patch $D$ and an antipodal causal patch $\overbar D$. The field's Hilbert space in $dS_4$ can be parameterized by wavefunctionals $\Psi[\varphi(x\in\Sigma)]$ on any global spatial slice $\Sigma$. Let us choose a $\Sigma$ that passes through the observer's bifurcation surface. The bifurcation surface then divides it into two halves $\Sigma_D$,$\Sigma_{\overbar D}$, which act as Cauchy slices for $D$ and $\overbar D$, respectively. We choose $\Sigma$ such that it is its own antipodal image, i.e. such that $\Sigma_D$ and $\Sigma_{\overbar D}$ are mutually antipodal. The simplest example of such a slice $\Sigma$ would be an equatorial 3-sphere in $dS_4$. In the basis of fields on $\Sigma$, the antipodal-map operator $\calA$ takes the form:
\begin{align}
 \calA\Psi[\varphi(x)] = \Psi^*[\eta \varphi(-x)] \ . \label{eq:A_Sigma}
\end{align}
Here, the complex conjugation flips the field's canonical momentum, thus realizing the time-reversal component of the antipodal map. Since the antipodal map is a symmetry of the theory, eq. \eqref{eq:A_Sigma} describes the same operator for all choices of $\Sigma$. Antipodally symmetric states in $dS_4$ are characterized by $\calA\Psi = \Psi$. Since $\calA$ is antiunitary, this is not a complex-linear equation, but a reality condition. Hence, the antipodally symmetric states do not form a Hilbert subspace, but instead form a \emph{real slice} of the full $dS_4$ Hilbert space. This is the reason why $dS_4/\bbZ_2$ doesn't admit an observer-independent quantization.

Back in ordinary $dS_4$, the antilinear operator $\calA$ can be equivalently viewed as a \emph{linear} map between ket vectors and bra vectors. When restricted to a single causal patch, it maps e.g. kets on $\overbar D$ to bras on $D$. This allows us to define a linear map between vectors $|\Psi\rangle$ in the global Hilbert space and operators $\hat\Psi$ on the observer's causal patch:
\begin{align}
 |\Psi\rangle = \sum_k c_k \left|\Psi^{(k)}_D\right> \left|\Psi^{(k)}_{\overbar D}\right> \quad \longleftrightarrow \quad 
 \hat\Psi = \sum_k c_k \left|\Psi^{(k)}_D\right> \left<\calA\Psi^{(k)}_{\overbar D}\right| \ .
 \label{eq:state_operator}
\end{align}
In the field configuration basis on the spatial slice $\Sigma$, the map \eqref{eq:state_operator} is simply:
\begin{align}
 \big\langle \varphi_D(x), \varphi_{\overbar D}(x) \big|\Psi\big\rangle = 
 \big\langle \varphi_D(x) \big| \hat\Psi \big| \eta \varphi_{\overbar D}(-x) \big\rangle \ . \label{eq:state_operator_configuration}
\end{align}
Here, $\varphi_D(x)$ and $\varphi_{\overbar D}(x)$ are field configurations on $\Sigma_D$ and $\Sigma_{\overbar D}$ respectively, while $\eta \varphi_{\overbar D}(-x)$ is a configuration on $\Sigma_D$.

Note that the global state $|\Psi\rangle$ is antipodally symmetric if and only if the operator $\hat\Psi$ is Hermitian. Also, the mixed state $\hat\rho_D$ which is the restriction of $|\Psi\rangle$ to the causal patch can be expressed as:
\begin{align}
 \hat\rho_D = \sum_{j,k} c_j c^*_k\, \big\langle\Psi^{(k)}_{\overbar D}\big|\Psi^{(j)}_{\overbar D}\big\rangle \left|\Psi^{(j)}_D\right> \left<\Psi^{(k)}_D\right| 
   = \hat\Psi\hat\Psi^\dagger \ . \label{eq:rho_psi}
\end{align}
Thus, the tracing out of the $\overbar D$ degrees of freedom is accomplished by the operator squaring operation $\hat\Psi\rightarrow\hat\Psi\hat\Psi^\dagger$ in the Hilbert space on $D$. The product with the opposite ordering, i.e. $\hat\Psi^\dagger\hat\Psi$, produces the state $\calA\,\hat\rho_{\overbar D}\,\calA^{-1}$, where $\hat\rho_{\overbar D}$ is the restriction of $|\Psi\rangle$ to $\overbar D$. In particular, we see that $\hat\rho_{\overbar D}$ is the antipodal image of $\hat\rho_D$ if and only if the operator $\hat\Psi$ is normal:
\begin{align}
 \hat\rho_{\overbar D} = \calA\,\hat\rho_D\calA^{-1} \quad \Longleftrightarrow \quad \hat\Psi\hat\Psi^\dagger = \hat\Psi^\dagger\hat\Psi \ . \label{eq:normal}
\end{align}
This is a weaker condition than Hermiticity $\hat\Psi^\dagger=\hat\Psi$, which encodes the antipodal symmetry of the entire global state $|\Psi\rangle$. When the condition \eqref{eq:normal} holds, the operator $\hat\Psi$ is unitarily diagonalizable. One can then pick the $|\Psi^{(k)}_D\rangle$ states in \eqref{eq:state_operator} to be orthonormal, with  $|\Psi^{(k)}_{\overbar D}\rangle$ their antipodal images. The expressions \eqref{eq:state_operator}-\eqref{eq:rho_psi} then simplify to:
\begin{gather}
 |\Psi\rangle = \sum_k c_k \left|\Psi^{(k)}_D\right> \left|\calA\Psi^{(k)}_D\right> \quad \longleftrightarrow \quad 
   \hat\Psi = \sum_k c_k \left|\Psi^{(k)}_D\right> \left<\Psi^{(k)}_D\right| \ ; \label{eq:Psi_normal} \\
 \hat\rho_D = \hat\Psi\hat\Psi^\dagger = \hat\Psi^\dagger\hat\Psi = \sum_k \left|c_k\right|^2 \left|\Psi^{(k)}_D\right> \left<\Psi^{(k)}_D\right| \ . \label{eq:rho_psi_normal}
\end{gather}

\subsubsection{Antipodally symmetric basis on the global slice} \label{sec:QFT:dS:antipodal_basis}

To facilitate the eventual application to $dS_4/\bbZ_2$, let us now express the map \eqref{eq:state_operator} in an antipodally symmetric basis on the global slice $\Sigma$. For antipodally even (odd) fields, such a basis can be constructed from the even (odd) part of configuration space $\varphi(x)$ together with the odd (even) part of momentum space $\dot \varphi(x)$:
\begin{align}
 \varphi_\text{sym}(x) = \frac{\varphi(x) + \eta \varphi(-x)}{2} \quad ; \quad \dot \varphi_\text{sym}(x) = \frac{\dot \varphi(x) - \eta\dot \varphi(-x)}{2} \ . \label{eq:phi_sym}
\end{align}
The antipodally symmetric fields $\varphi_\text{sym},\dot \varphi_\text{sym}$ on the global slice $\Sigma$ can be parameterized by their values on the half-slice $\Sigma_D$. Note that $\varphi_\text{sym}$ and $\dot \varphi_\text{sym}$ commute, even though $\varphi$ and $\dot \varphi$ do not. In a free field theory, eq. \eqref{eq:phi_sym} defines the same basis for all choices of $\Sigma$. This isn't true in general for interacting theories, since the non-linear evolution will mix the antipodally even and odd components of the field. 

The antipodal map in the basis \eqref{eq:phi_sym} is just a complex conjugation:
\begin{align}
 \calA\Psi[\varphi_\text{sym}(x), \dot \varphi_\text{sym}(x)] = \Psi^*[\varphi_\text{sym}(x), \dot \varphi_\text{sym}(x)] \ . \label{eq:A_antipodal} 
\end{align}
Thus, antipodally symmetric states have real wavefunctionals in the antipodally symmetric basis.

The transformation between the configuration basis $|\varphi\rangle$ and the antipodally symmetric basis $|\varphi_\text{sym},\dot \varphi_\text{sym}\rangle$ is a Fourier transform between $\varphi(x) - \eta \varphi(-x) \equiv 2\varphi_\text{skew}(x)$ and its canonical conjugate, $\dot\varphi_\text{sym}(x)$. As a result, the map \eqref{eq:state_operator_configuration} between states in the global Hilbert space and operators in the causal patch becomes:
\begin{align}
 \begin{split}
   &\big\langle \varphi_\text{sym}(x), \dot \varphi_\text{sym}(x) \big|\Psi\big\rangle \\
   &\quad  = \int\calD\varphi_\text{skew}\,e^{-2i\dot\varphi_\text{sym}\cdot\varphi_\text{skew}}\, 
         \big\langle \varphi_\text{sym}(x) + \varphi_\text{skew}(x) \big| \hat\Psi \big| \varphi_\text{sym}(x) - \varphi_\text{skew}(x) \big\rangle \ .
 \end{split} \label{eq:state_operator_antipodal}
\end{align}
Here, $x$ refers to points on $\Sigma_D$, and the matrix element on the RHS is between two field configurations $\varphi_\text{sym}(x) \pm \varphi_\text{skew}(x)$ on $\Sigma_D$. The scalar product $\dot\varphi_\text{sym}\cdot\varphi_\text{skew}$ stands for:
\begin{align}
 \dot\varphi_\text{sym}\cdot\varphi_\text{skew} \equiv \int_{\Sigma_D} d^3x\, \dot\varphi_\text{sym}(x)\, \varphi_\text{skew}(x) \ .
\end{align}
Note that eq. \eqref{eq:state_operator_antipodal} has the form of a Wigner-Weyl transform between a phase space functional on the LHS and a quantum operator on the RHS:
\begin{align}
 f(q,p) = \int dq'\, e^{-ipq'}\, \big\langle q + \frac{q'}{2} \big| \hat f \big| q - \frac{q'}{2} \big\rangle \ . \label{eq:WW}
\end{align}
The difference is that the variables $\varphi_\text{sym},\dot \varphi_\text{sym}$ refer not to the phase space on $\Sigma_D$, but to the antipodally symmetrized combinations \eqref{eq:phi_sym} on $\Sigma$. This distinction will disappear once we switch perspective from $dS_4$ to $dS_4/\bbZ_2$. As an aside, we note that a similar structure has appeared in the context of string field interactions \cite{Bars:2014vca,Bars:2014jca}.

So far, we've been parameterizing the Hilbert space in terms of a bulk spatial slice $\Sigma$ that passes through the observer's bifurcation surface. From the point of view of holography, it is more natural to use one of the horizons $H_i,H_f$, which can be viewed as limiting cases. Working on a null horizon simplifies some technical aspects, while complicating others. The horizon version of the map \eqref{eq:state_operator_antipodal} is presented in Appendix \ref{app:horizon_map}. It is a straightforward adaptation of tools developed by the authors in \cite{Halpern:2015cia}.

\subsubsection{Example: Bunch-Davies meta-state and thermal states} \label{sec:QFT:dS:BD}

An important example of the map \eqref{eq:state_operator} is when the global state $|\Psi\rangle$ is the Bunch-Davies vacuum \cite{Bunch:1978yq}. This special case will be instrumental for the holographic construction in section \ref{sec:dS/CFT}. As we will show below, the corresponding operator $\hat\Psi$ on the causal-patch Hilbert space reads:
\begin{align}
 \hat\Psi_0 = \hat R\, e^{-\pi\hat H} \ . \label{eq:Psi_BD}
\end{align}
Here, $\hat H$ is the Hamiltonian in the static coordinates \eqref{eq:static}, i.e. the generator of evolution along the Killing vector $\xi^\mu$ from figure \ref{fig:time}. $\hat R$ is the operator that implements the antipodal map on the $(\theta,\phi)$ 2-spheres in static coordinates. This 2d antipodal map is the non-trivial central element in the observer's residual symmetry group $\bbR\times O(3)$. It is a discrete symmetry of P type in the CPT classification. Thus, the operator $\hat R$ exists in any P-invariant theory, and therefore in any theory that respects the full $dS_4$ antipodal map. 

The restriction of the Bunch-Davies vacuum into the causal patch is of course the thermal state at the de Sitter temperature, which can be derived from \eqref{eq:Psi_BD} as:
\begin{align}
 \hat\rho_D = \hat\Psi_0 \hat\Psi_0^\dagger = e^{-2\pi\hat H} \ . \label{eq:rho_thermal}
\end{align}

We will now sketch three derivations of eq. \eqref{eq:Psi_BD}. The first is similar to the Euclidean path integral derivation of the Unruh effect. Consider an equatorial $S_3$ slice $\Sigma$ in $dS_4$, divided in half by the observer's bifurcation surface into Cauchy slices $\Sigma_D,\Sigma_{\overbar D}$ for the causal patch and its antipode. The Bunch-Davies vacuum wavefunctional on $\Sigma$ can be viewed as the output of a Euclidean path integral over a half-$S_4$. This same path integral can also be viewed as an operator mapping bras on $\Sigma_{\overbar D}$ to kets on $\Sigma_D$, via a boost by an imaginary angle $-\pi i$ in the plane $p_i\wedge p_f$ in $\bbR^{1,4}$. On the other hand, note that boosts in the $p_i\wedge p_f$ plane are just time translations in the observer's static coordinates. The result \eqref{eq:Psi_BD} then follows, after using the spacetime antipodal map to further reinterpret the path integral as an operator from kets to kets on $\Sigma_D$. 

Alternatively, instead of using a spatial $S_3$ slice, one can derive \eqref{eq:Psi_BD} on one of the observer's horizons $H_i,H_f$. Let $u$ be an affine null coordinate along the horizon's generators, such that the bifurcation surface is at $u=0$. Through the magic of lightfront quantization \cite{Leutwyler:1970wn}, the Bunch-Davies vacuum can be defined \emph{kinematically} on the horizon, as the lowest eigenstate of the generator of $u$ translations \cite{Gibbons:1977mu}. The corresponding operator \eqref{eq:Psi_BD} can then be calculated explicitly, by slightly modifying a similar calculation in \cite{Halpern:2015cia}. 

The third and final derivation method combines the advantages of the previous two. The idea is to express the Bunch-Davies wavefunctional on a horizon, e.g. $H_f$, as the result of a ``path integral'' over the upper half of the complex $u$ plane. This is similar to the Euclidean path integral definition of the Minkowski vacuum; however, instead of evolving a spatial slice along a transverse time direction, here one is ``evolving'' the real $u$ axis in the direction of imaginary $u$, using the \emph{kinematical} generator of $u$ translations. Thus, this horizon version of the ``Euclidean path integral'' lives on a complexification of the horizon, rather than of the whole spacetime. Now, similarly to the $S_3$-based argument, we can reinterpret this ``path integral'' as a \emph{phase rotation} by $\pi$ in the complex $u$ plane, mapping the real half-axis $u<0$ (which causally spans $\overbar D$) onto the real half-axis $u>0$ (which causally spans $D$). This phase rotation can be expressed as a translation by $-\pi i$ in the null coordinate $\tau=\ln u$. However, $\tau$ translations on the horizon are equivalent to time translations in the causal patch, since both are generated by the same Killing vector $\xi^\mu$. Eq. \eqref{eq:Psi_BD} now follows as in the $S_3$-based argument, after relating the two half-horizons via the spacetime antipodal map.

\subsection{Quantum fields in $dS_4/\bbZ_2$} \label{sec:QFT:elliptic}

We now turn to quantum field theory in $dS_4/\bbZ_2$, making use of the tools developed for $dS_4$ in section \ref{sec:QFT:dS}. Roughly speaking, we will identify quantum states in $dS_4/\bbZ_2$ with antipodally symmetric states in the $dS_4$ double cover, i.e. states that satisfy $\calA\Psi=\Psi$. However, the precise procedure will be more subtle, since the antipodally symmetric states in $dS_4$ form a real slice, not a complex subspace, of the $dS_4$ Hilbert space. Thus, they cannot be identified directly with a $dS_4/\bbZ_2$ Hilbert space, which in fact can only be defined in the context of an observer \cite{Hackl:2014txa}. When discussing $dS_4/\bbZ_2$, we will refer to antipodally symmetric $dS_4$ states as ``meta-states'' -- a term reflecting their observer-independence and their non-trivial relation to actual states in $dS_4/\bbZ_2$. As will become clear, these ``meta-states'' are the same as those introduced in \cite{Hackl:2014txa}.

\subsubsection{Causal patches in $dS_4/\bbZ_2$ and their embedding in $dS_4$}

A priori, there is no quantum theory in $dS_4/\bbZ_2$, since the lack of time-orientability makes the signs of commutators ambiguous. An operator algebra and Hilbert space appear only in relation to an observer (which we define, as before, by the initial and final worldline endpoints $p_i,p_f\in\scri$). Recall that an observer's causal patch in $dS_4/\bbZ_2$ carries a time orientation, induced by the ordering of $p_i,p_f$. Therefore, quantum field theory inside the causal patch can be done as usual. This is all that is needed to have consistent physics for the observer, since only the inside of the causal patch is operationally accessible (see sections \ref{sec:geometry:observer},\ref{sec:geometry:elliptic_observers}). Each observer defines an operator algebra and Hilbert space on a different causal patch in $dS_4/\bbZ_2$. 

The causal patch of an observer in $dS_4/\bbZ_2$ has a natural image in the double-cover space $dS_4$. In general, a point in $dS_4/\bbZ_2$ has two antipodally related images in $dS_4$. Thus, a causal patch $D$ becomes a pair of antipodally related causal patches in $dS_4$. However, the ordering of the worldline endpoints $p_i,p_f$ allows us to say which of these two patches is ``really'' $D$, and which is the antipodal patch $\overbar D$. Indeed, out of the two $dS_4$ images of $p_i\in\scri$, we can unambiguously identify the one on $\scri^-$ as the ``true'' image, and the one on $\scri^+$ as the antipode. Similarly, the ``true'' image of $p_f\in\scri$ will be the one on $\scri^+$. Having thus distinguished the ``true'' images $p_i,p_f$ from their antipodes $\bar p_i,\bar p_f$, we can identify the image of $D$ in $dS_4$ as the causal patch with endpoints $p_i,p_f$, as opposed to its antipode $\overbar D$ with endpoints $\bar p_f,\bar p_i$. 

Consider now two different observers in $dS_4/\bbZ_2$. The spacetime region accessible to both is the overlap of their causal patches $D,D'$. This is also the region where both observers assign a time orientation. In general, it will consist of two subregions: one where the observers agree on the time orientation (and thus on the operator algebra), and one where they disagree. The $dS_4$ image of the region where the observers assign the same time orientation is $D\cap D'$, while the image of the region where they assign opposite orientations is either $D\cap\overbar D'$ or $D'\cap \overbar D$, depending on which observer's perspective is taken as primary.

\subsubsection{From meta-states to states}

We are now ready to describe how an antipodally symmetric ``meta-state'' $\Psi$ in $dS_4$ translates into states viewed by observers in $dS_4/\bbZ_2$. The idea is simple: we embed the observer's causal patch $D$ into $dS_4$ as described above, and then find the restriction $\hat\rho_D$ of the $dS_4$ state $\Psi$ to $D$. We interpret this restricted state $\hat\rho_D$ as the density matrix for the state in $dS_4/\bbZ_2$ seen by the observer; note that, in general, this state will be mixed. Thus, the Hilbert space of the observer in $dS_4/\bbZ_2$ is identified with the ordinary Hilbert space on the image of $D$ in $dS_4$. Another observer, with causal patch $D'$, will have a different Hilbert space, on which the meta-state $\Psi$ will induce a different mixed state $\hat\rho_{D'}$. In particular, the mixed states seen by different observers may have different entropies, as we will see in an example below.

It is useful to think of the restriction $\Psi\rightarrow\hat\rho_D$ in two steps, described by eqs. \eqref{eq:Psi_normal}-\eqref{eq:rho_psi_normal}. The first step is to translate the global $dS_4$ state $\Psi$ into a Hermitian operator $\hat\Psi$ on the Hilbert space in $D$. Since $\Psi$ is antipodally symmetric, $\hat\Psi$ will be Hermitian. The second step is to calculate the restriction to $D$ as $\hat\rho_D = \hat\Psi\hat\Psi^\dagger = \hat\Psi^2$. Note that the first step $\Psi\rightarrow\hat\Psi$ involves no information loss, and can be reversed. However, in the second step $\hat\Psi\rightarrow\hat\rho_D$, we lose the information that was captured in the entanglement between $D$ and $\overbar D$ in the global $dS_4$ state $\Psi$. In our present formalism, this lost information is encoded in the signs of the eigenvalues of $\hat\Psi$.

If the $dS_4$ wavefunctional $\Psi$ happens to be written in an antipodally symmetric basis on a spatial or null slice as in section \ref{sec:QFT:dS:antipodal_basis}, then one can reinterpret it as a \emph{functional on the phase space} on the corresponding slice in $dS_4/\bbZ_2$. This phase space is shared by all observers whose bifurcation surface lies on the chosen slice; however, each observer will endow it with a different symplectic structure, arising from her notion of time orientation. In this setup, the map $\Psi\rightarrow\hat\Psi$ becomes a Wigner-Weyl transform, as we've seen in eqs. \eqref{eq:state_operator_antipodal}-\eqref{eq:WW} or \eqref{eq:state_operator_null}-\eqref{eq:WW_momentum}. This establishes that our recipe for deriving states from meta-states is the same as that in \cite{Hackl:2014txa}, except for the choice of slice on which the phase space is parameterized.

\subsubsection{Example: Bunch-Davies meta-state and thermal states}

A particular example studied in \cite{Hackl:2014txa} was to choose the Bunch-Davies vacuum in $dS_4$ as the meta-state $\Psi$. As we've seen in section \ref{sec:QFT:dS:BD}, the corresponding operator $\hat\Psi$ on an observer's Hilbert space is given by \eqref{eq:Psi_BD}, leading to the thermal state \eqref{eq:rho_thermal} in a causal patch. In \cite{Hackl:2014txa}, these results were derived by brute force for a particular field theory. Here, we were able to derive them generally, having understood the physical meaning of the map from meta-states to states.

The thermal state \eqref{eq:rho_thermal} appears the same to all observers, since it arises from a meta-state that is invariant under the de Sitter group. It was pointed out in \cite{Hackl:2014txa} that using a more general $\alpha$-vacuum \cite{Allen:1985ux} instead of the Bunch-Davies vacuum leads to singular states for the $dS_4/\bbZ_2$ observers. In our present framework, it becomes clear that this is just the well-known singularity of $\alpha$-vacua restricted to the static patch.

\subsubsection{Translating states between observers}

We have seen how, starting from a meta-state, one can work out the state for each observer. However, the meta-state itself is of course not accessible to the observers. Thus, a more realistic scenario is to start from the state according to some observer, and then try to work out which states will be seen by others. For classical field solutions, as opposed to quantum states, such a task would be straightforward: each observer's causal patch causally spans $dS_4/\bbZ_2$, and thus the contents of other patches can be determined by evolution. At the quantum level, the situation is more subtle. Given a general mixed state $\hat\rho_D$ on her Hilbert space, an observer must guess the meta-state $\Psi$ before she can deduce the state seen by other observers. This involves reversing the squaring operation $\hat\Psi\rightarrow \hat\rho_D = \hat\Psi^2$, which can only be done up to a sign ambiguity for each nonzero eigenvalue. 

One possibility is to let our observer sum over all sign choices with equal probabilities, producing a \emph{mixed} state in $dS_4$ as the meta-state. It is easy to see from eqs. \eqref{eq:Psi_normal}-\eqref{eq:rho_psi_normal} that this mixed meta-state is just the non-entangled product $\hat\rho_D\otimes\calA\hat\rho_D\calA^{-1}$ of $\hat\rho_D$ with its antipodal image. The mixed meta-state can then be used as before to induce mixed states in other causal patches. Due to the non-entangled structure of the meta-state, it contains a firewall at the original observer's bifurcation surface, which either lies inside or can propagate into the causal patches of other observers. However, one must remember that this firewall-containing mixed meta-state is no more than an assessment of ignorance by the original observer. Her mixed state $\hat\rho_D$ may in fact be arising from a perfectly regular meta-state, as in the Bunch-Davies/thermal example above.

One situation where our observer can unambiguously deduce the state seen by another is when the two observers have the same worldline endpoints $p_i,p_f$, but in reversed order, i.e. the second observer has endpoints $(p_i',p_f')=(p_f,p_i)$. Such a pair of observers share the same causal patch in $dS_4/\bbZ_2$. However, in the $dS_4$ double cover, they are associated with antipodal causal patches, i.e. $D' = \overbar D$. Thus, if the first observer sees a mixed state $\hat\rho_D$, the antipodal symmetry of the meta-state implies that the second observer will see its antipodal image $\hat\rho_{D'} = \calA\hat\rho_D\calA^{-1}$. Note that these two states do not share a Hilbert space, since they live on different causal patches in $dS_4$.

Another special case is when the original observer has full knowledge of the state in her causal patch, i.e. when she sees a pure state $\hat\rho_D = |\Psi_D\rangle\langle\Psi_D|$. Then the squaring operation $\hat\Psi\rightarrow\hat\rho_D$ can be reversed unambiguously as $\hat\Psi = \hat\rho_D$, up to an irrelevant \emph{overall} sign. The meta-state is then the non-entangled product $|\Psi_D\rangle|\calA\Psi_D\rangle$, which induces firewall-containing mixed states in the causal patches of other observers. The state is pure only for the original observer and for the ``antipodal'' observer with $(p_i',p_f')=(p_f,p_i)$. This is contrary to the claim \cite{Parikh:2002py} that the states seen by different observers are always related by Bogolyubov transformations.

\subsubsection{Example: two observers that share a horizon or Cauchy slice} \label{sec:QFT:elliptic:shared_slice}

Let us explore in more detail how one observer's pure state induces a mixed state for another observer. Consider a pair of observers in $dS_4/\bbZ_2$ with causal patches $D,D'$, such that their bifurcation surfaces lie on a shared spacelike hypersurface. In the limit in which the shared hypersurface is a null horizon, the observers share one of their worldline endpoints.

Let us denote by $\Sigma$ the $dS_4$ image of the spacelike or null hypersurface in $dS_4/\bbZ_2$ that contains both bifurcation surfaces. The bifurcation surfaces divide $\Sigma$ into four pieces $\Sigma_{D\cap D'}$,$\Sigma_{D\cap\overbar D'}$,$\Sigma_{\overbar D\cap D'}$,$\Sigma_{\overbar D\cap\overbar D'}$, lying in the intersections of the corresponding causal patches. $\Sigma_{D\cap D'}$ and $\Sigma_{\overbar D\cap\overbar D'}$ are mutually antipodal, and correspond in $dS_4/\bbZ_2$ to a region where the two observers induce the same time orientation (let us call this ``region $I$''). $\Sigma_{D\cap\overbar D'}$ and $\Sigma_{\overbar D\cap D'}$ are again mutually antipodal, and correspond in $dS_4/\bbZ_2$ to a region where the observers induce \emph{opposite} time orientations (let us call this ``region $II$''). 

When $\Sigma$ is spacelike, the full Hilbert space on it (i.e. the Hilbert space on $dS_4$) is a direct product of the Hilbert spaces on each piece. When $\Sigma$ is a null horizon, this becomes more subtle, since points on the same lightray are causally connected. However, as discussed in \cite{Halpern:2015cia}, the subtlety is confined to modes of zero frequency in the null coordinate. Disregarding such modes, which are associated with a host of issues in lightfront quantization \cite{Maskawa:1975ky,Yamawaki:1997cj}, we can again express the Hilbert space on $\Sigma$ as a product of the Hilbert spaces on each piece.

With these preliminaries in place, consider a pure state $\Psi_D$ on $\Sigma_D$, i.e. in the first observer's Hilbert space. As discussed above, the corresponding meta-state is unambiguous, and is given by a product $|\Psi_D\rangle|\calA\Psi_D\rangle$ of a state on $D$ and a state on $\overbar D$. The restriction of this meta-state to $D'$, i.e. the mixed state seen by the second observer, is then given by a non-entangled product of a state on $D\cap D'$ and a state on $\overbar D\cap D'$:
\begin{align}
 \hat\rho_{D'} = \hat\rho_{D\cap D'}\otimes\calA\,\hat\rho_{D\cap\overbar D'}\,\calA^{-1} \ , \label{eq:non_entangled}
\end{align}
where $\hat\rho_{D\cap D'}$ and $\hat\rho_{D\cap\overbar D'}$ are the restrictions of the first observer's state to $\Sigma_{D\cap D'}$ and $\Sigma_{D\cap\overbar D'}$, respectively. Intuitively, the situation can be described as follows (see figure \ref{fig:pure_to_mixed}):
\begin{figure}%
\centering%
\includegraphics[scale=.8]{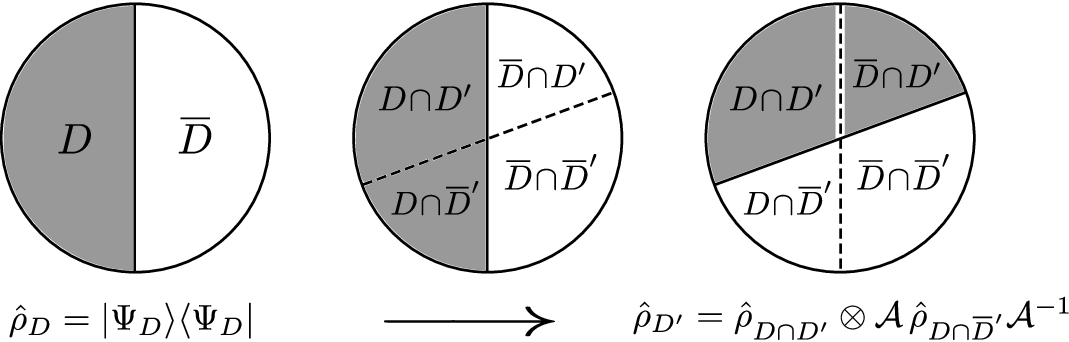} \\
\caption{This figure depicts how a pure state for one observer in $dS_4/\bbZ_2$ translates into a mixed state for another observer. The disks represent a spatial slice of the $dS_4$ double cover which contains the two observers' bifurcation surfaces (represented by straight lines). $D$ and $D'$ denote the observers' causal patches, while $\overbar D$ and $\overbar D'$ are their antipodes in $dS_4$.}
\label{fig:pure_to_mixed} 
\end{figure}%
\begin{enumerate}
 \item We find the restrictions $\hat\rho_{D\cap D'},\hat\rho_{D\cap\overbar D'}$ of the first observer's state on $\Sigma_D$ to the subregions $\Sigma_{D\cap D'},\Sigma_{D\cap\overbar D'}$.
 \item On $\Sigma_{D\cap D'}$, the observers agree on time orientation. The state there can be left as is.
 \item On $\Sigma_{D\cap\overbar D'}$, the observers disagree on time orientation. One must therefore reverse it by applying the antipodal map.
 \item Since the time-reversing antipodal map is anti-unitary, we lose the interference phases that encoded the entanglement between the two subregions. Thus, the state seen by the second observer is the non-entangled product \eqref{eq:non_entangled}.
\end{enumerate}
The non-entangled product state \eqref{eq:non_entangled} seen by the second observer contains a firewall along the first observer's bifurcation surface. 
The entropy of the state \eqref{eq:non_entangled} can be expressed as:
\begin{align}
 \sigma[\hat\rho_{D'}] = \sigma[\hat\rho_{D\cap D'}] + \sigma[\hat\rho_{D\cap\overbar D'}] = 2\sigma_\text{ent} \ ,
\end{align}
where $\sigma_\text{ent}$ is the entanglement entropy of the first observer's state across the second observer's bifurcation surface. 

To conclude, a pure state seen by one observer cannot be translated into another observer's world-picture without a loss of information. This loss arises from a decoherence between two regions whose relative time orientation changes. The lost information is directly related to the entanglement entropy across the observers' horizons.

\subsubsection{Summary: What different observers see in different regions of $dS_4/\bbZ_2$} \label{sec:QFT:elliptic:summary}

Let us summarize what we've learned about the states assigned by observers to different regions of $dS_4/\bbZ_2$. In general, quantum field theory in $dS_4/\bbZ_2$ only makes sense in an observer's causal patch. Different observers have different quantum world-pictures with different operator algebras and Hilbert spaces. However, states in the various causal patches can be thought of as arising from a shared ``meta-state'' in the $dS_4$ double cover. A causal patch in $dS_4/\bbZ_2$ has a unique embedding into $dS_4$ thanks to the time orientation induced by the observer. This is not the case for the region \emph{outside} the causal patch. It is thus unclear in general how to incorporate this region into the observer's world-picture. However, this doesn't mean that the bulk physics is incomplete, since the region outside the causal patch is operationally irrelevant (see sections \ref{sec:geometry:observer},\ref{sec:geometry:elliptic_observers}). 

We can compare the observations of two observers in the overlap of their causal patches $D,D'$. This overlap can be divided into ``region $I$'', where the observers agree on the time orientation, and ``region $II$'', where they disagree. In region $I$, the observers will agree on the state, while in region $II$, their states will be related through the antipodal map $\calA$. The observers will in general disagree on correlations \emph{between} regions $I$ and $II$. For instance, in the setup of section \ref{sec:QFT:elliptic:shared_slice}, the first observer sees entanglement between the two regions, while the second observer doesn't.

\subsubsection{An aside: Translating observables between observers}

In \cite{Hackl:2014txa}, apart from the recipe for translating states, an additional recipe was proposed for translating \emph{observable operators} between observers. This observable-translation recipe uses the map \eqref{eq:state_operator} that relates operators on the causal patch to Hilbert-space vectors in $dS_4$ (or, in the language of \cite{Hackl:2014txa}, to functionals on the $dS_4/\bbZ_2$ phase space). This creates a dictionary between operators on different causal patches, via mapping into the shared $dS_4$ Hilbert space and back again. The procedure is identical to our translation recipe for states, except that the squaring operation $\hat\Psi\rightarrow\hat\rho_D = \hat\Psi^2$ has been omitted. 

It was noted in \cite{Hackl:2014txa} that this observable-translation recipe does not in general lead to agreement on expectation values, except in the classical limit. From our present perspective, the recipe in general doesn't have a clear physical meaning at all. However, for operators localized either in region $I$, where two observers agree on the time orientation, or in region $II$, where they disagree, it's easy to show that the recipe of \cite{Hackl:2014txa} reduces to the map $\hat\calO\leftrightarrow\hat\calO$ or $\hat\calO\leftrightarrow\calA\hat\calO\calA^{-1}$ in the $dS_4$ double cover, respectively. The observers then \emph{do} agree on the operator's expectation value, up to a complex conjugation in the second case.

\section{dS/CFT for free quantum fields in a causal patch} \label{sec:dS/CFT}

In this section, we present the central result of the paper. Using the ingredients from previous sections, we construct holographically the operator algebra and Hamiltonian of free massless fields for an observer in $dS_4/\bbZ_2$. Since free massless fields form the linearized limit of higher-spin gravity, we view this as a first step towards extracting causal-patch physics from the full higher-spin dS/CFT. The discussion is structured as follows. In section \ref{sec:dS/CFT:problem}, we make precise what we mean by a holographic description of the free-field operator algebra and Hamiltonian in the causal patch. In section \ref{sec:dS/CFT:solution}, we show how this can be achieved in the massless case by combining the standard dS/CFT of \cite{Maldacena:2002vr,Anninos:2011ui} with the boundary encoding of antipodal symmetry from section \ref{sec:encoding:free}. Finally, in section \ref{sec:dS/CFT:S_matrix}, we discuss how the same method does \emph{not} capture the ``S-matrix'' between the observer's past and future horizons.

\subsection{Free-field holography in $dS_4/\bbZ_2$: the problem statement} \label{sec:dS/CFT:problem}

In section \ref{sec:QFT:elliptic}, we've seen how to make sense of quantum fields in $dS_4/\bbZ_2$ \emph{inside an observer's causal patch}. As discussed in section \ref{sec:geometry:observer}, this is all that is needed to predict the outcomes of experiments in this spacetime. However, our long-term goal is not quantum field theory in $dS_4/\bbZ_2$, but quantum gravity. The lesson from AdS/CFT is that our best hope for defining such a theory is in terms of the spacetime's boundary $\scri$. With this motivation, let us see how a non-gravitational QFT in the causal patch can be expressed in terms of $\scri$. In classical theory, this is a trivial matter of evolving with the field equations; indeed, our initial motivation for working in $dS_4/\bbZ_2$ is that $\scri$ is causally determined by an observer's causal patch. However, at the quantum level, subtleties arise. Recall that an observer induces a time orientation, and with it a standard operator algebra, only inside her causal patch. This makes it hard to define evolution between the causal patch and $\scri$, which only intersects the patch at the two endpoints $(p_i,p_f)$. Our temporary solution is to focus on \emph{free} field theory, where this problem disappears: the evolution becomes independent of the operator algebra, since it is linear in the field operators. At the interacting level, our hope is that  higher-spin gravity, due to its special structure, will find its own way around the problem (see also section \ref{sec:discuss}).

In free field theory on $dS_4/\bbZ_2$, the classical field equations determine the fields everywhere as linear functionals of fields on a Cauchy slice inside the causal patch (or on one of the horizons, as a limiting case). As discussed above, these evolution equations translate trivially to the quantum level, despite the absence of a standard operator algebra outside the patch: products of field operators simply never arise. With evolution out of the causal patch defined in this way, it \emph{induces} an operator algebra in the outside region: fields outside the patch can be expressed as linear combinations of fields inside, for which we have an algebra. However, the algebra induced in this way on the outside region is not the standard one. Indeed, this must be the case, since the outside region is not time-oriented: there is no natural way to identify e.g. the direction towards $\scri$ as either future or past. The closest thing to a time orientation is given by the observer's time-translation Killing field $\xi^\mu$ (see figure \ref{fig:time}), which becomes spacelike outside the causal patch. Indeed, we will see that spacelike-separated field operators in the outside region (and in particular on $\scri$) do not commute.

With these preliminaries in place, we are ready to pose the task of ``free-field holography'' in $dS_4/\bbZ_2$. It is to find the algebra that an observer induces on $\scri$, and to identify within it the Hamiltonian operator that generates translations along $\xi^\mu$. This will mean, by construction, that we've expressed the operator algebra and Hamiltonian in the observer's causal patch in terms of asymptotic quantities on $\scri$. Knowing the operator algebra, one can also reconstruct the observer's Hilbert space as a Fock space generated by creation and annihilation operators on $\scri$.

Let us now pose the task more explicitly, using what we know about free fields in $dS_4/\bbZ_2$. For simplicity, we again consider a spin-0 field $\varphi$, which can be either an antipodally even scalar or an antipodally odd pseudoscalar. The discussion below will extend straightforwardly to nonzero integer spins. For now, we keep the field's mass arbitrary. Let us consider the asymptotic boundary data $\varphi_\Delta$ on $\scri$, with one of the conformal weights $\Delta = \Delta_\pm$. In the massless case, we choose the weight for which the boundary data remains unconstrained under the chosen antipodal symmetry. Then, in both the massive and massless cases, the data with the other weight is fixed by $\varphi_\Delta$ and the antipodal symmetry. Thus, our single boundary data $\varphi_\Delta$ spans the full solution space in $dS_4/\bbZ_2$, or, equivalently, the phase space in the observer's causal patch. 

To parameterize $\varphi_\Delta$, we will use a conformal frame and coordinates on $\scri$ adapted to the observer's residual $\bbR\times O(3)$ symmetry. First, let us choose a frame in the $\bbR^{1,4}$ embedding space such that the observer's worldline endpoints $p_i,p_f$ correspond to the following null directions:
\begin{align}
 p_i^\mu = (-1,1,\vec 0) \cong (1,-1,\vec 0) \quad ; \quad p_f^\mu = (1,1,\vec 0) \cong (-1,-1,\vec 0) \ . \label{eq:p_if}
\end{align}
In this frame, we represent $\scri$ as the following section of the (e.g. future) lightcone in $\bbR^{1,4}$:
\begin{align}
 \ell^\mu = (\cosh t, \sinh t, \vec n) \ , \label{eq:boundary_coords}
\end{align}
where $t$ is a coordinate with range $-\infty<t<\infty$, and $\vec n$ is a unit 3d vector parameterizing the 2-sphere. The lightcone section \eqref{eq:boundary_coords} has the topology and metric of a cylinder $\bbR\times S_2$, obtained from the conformal 3-sphere $\scri$ by removing the endpoints $p_i,p_f$ (which correspond to $t=\mp\infty$). The Killing vector $\xi^\mu$ that generates time translations in the causal patch acts on $\scri$ as translations in the (spacelike) $t$ coordinate. Similarly, the generators of $O(3)$ rotations in the causal patch act on $\scri$ by rotating the 2-spheres parameterized by $\vec n$. We can now expand the boundary data $\varphi_\Delta$ in modes parameterized by frequency $\omega$ and angular momentum quantum numbers $(l,m)$:
\begin{align}
 \varphi_\Delta(\cosh t, \sinh t,\vec n) = \int_0^\infty\frac{d\omega}{2\pi}\sum_{lm}\left(e^{-i\omega t}\, Y_{lm}(\vec n)\, c_{lm}(\omega) + c.c.\right) \ , \label{eq:boundary_modes}
\end{align}
where $c_{lm}(\omega),c_{lm}^*(\omega)$ are the mode coefficients, and $Y_{lm}(\vec n)$ are spherical harmonics.

Now, since the causal patch contains the same Killing vectors, we can decompose the bulk field $\varphi$ in the patch into harmonic oscillators with the same $(\omega,l,m)$ quantum numbers. These can be associated with annihilation and creation operators $\hat a_{lm}(\omega),\hat a^\dagger_{lm}(\omega)$, satisfying the standard commutation relations:
\begin{align}
 \begin{split}
   &\left[\hat a_{lm}(\omega), \hat a^\dagger_{l'm'}(\omega') \right] = 2\pi\delta(\omega-\omega')\delta_{ll'}\delta_{mm'} \ ; \\
   &\left[\hat a_{lm}(\omega), \hat a_{l'm'}(\omega') \right] = \left[\hat a^\dagger_{lm}(\omega), \hat a^\dagger_{l'm'}(\omega') \right] = 0 \ . 
 \end{split} \label{eq:a_algebra}
\end{align}
Note that by construction, $\omega$ is the time frequency of the oscillators. Thus, we can immediately express the Hamiltonian in the causal patch as:
\begin{align}
 \hat H = \int_0^\infty\frac{d\omega}{2\pi}\sum_{lm} \omega\, \hat a^\dagger_{lm}(\omega)\, \hat a_{lm}(\omega) \ . \label{eq:a_Hamiltonian}
\end{align}

The annihilation operators $a_{lm}(\omega)$ can be expressed explicitly in terms of the field $\varphi$ in the causal patch. Through evolution in $dS_4/\bbZ_2$, they can also be expressed in terms of the boundary data \eqref{eq:boundary_modes} on $\scri$. In fact, from the $\bbR\times O(3)$ symmetry and the linearity of free-field evolution, it is clear that $a_{lm}(\omega)$ must be the same as the boundary mode $c_{lm}(\omega)$, up to a complex normalization factor $N_l(\omega)$:
\begin{align}
 a_{lm}(\omega) = N_l(\omega)c_{lm}(\omega) \ . \label{eq:a_c}
\end{align}
Thus, in terms of the boundary modes, the operator algebra \eqref{eq:a_algebra} and the Hamiltonian \eqref{eq:a_Hamiltonian} take the form:
\begin{align}
 \begin{split}
   &\left[\hat c_{lm}(\omega), \hat c^\dagger_{l'm'}(\omega') \right] = \frac{1}{\left|N_l(\omega)\right|^2}\cdot 2\pi\delta(\omega-\omega')\delta_{ll'}\delta_{mm'} \ ; \\
   &\left[\hat c_{lm}(\omega), \hat c_{l'm'}(\omega') \right] = \left[\hat c^\dagger_{lm}(\omega), \hat c^\dagger_{l'm'}(\omega') \right] = 0 \ .
 \end{split} \label{eq:c_algebra} \\
 &\hat H = \int_0^\infty\frac{d\omega}{2\pi}\sum_{lm} \omega \left|N_l(\omega)\right|^2 \hat c^\dagger_{lm}(\omega)\, \hat c_{lm}(\omega) \ . \label{eq:c_Hamiltonian}
\end{align}
We conclude that in order to express the observer's operator algebra and Hamiltonian in terms of $\scri$, it suffices to find the absolute values $\left|N_l(\omega)\right|^2$ of the normalization factors $N_l(\omega)$. Of course, since we are dealing with free theory, these factors can be found explicitly e.g. by propagating the field from the boundary to one of the horizons $H_i,H_f$. For example, for an antipodally even conformally-coupled massless scalar, the answer reads \cite{Hackl:2014txa}:
\begin{align}
 \left|N_l(\omega)\right|^2 = \frac{1}{2\omega}\prod_{k=0}^l \left(\omega^2 + k^2 \right)^{(-1)^{l+k+1}} \ .
\end{align}
However, such an explicit calculation would defeat the purpose of the exercise. We instead wish to deduce $\left|N_l(\omega)\right|^2$ ``holographically'', without any bulk calculations. In the next subsection, we accomplish this for the massless case. 

\subsection{Solution for free massless fields using dS/CFT} \label{sec:dS/CFT:solution}

We now specialize to free \emph{massless} fields, which form the linearized limit of higher-spin theory. We again restrict for simplicity to a spin-0 field $\varphi$. We take it to be conformally-coupled massless, which is the relevant case for higher-spin theory. Again, the field can be either an antipodally even scalar or an antipodally odd pseudoscalar; the two cases are relevant respectively for the type-A and type-B versions of higher-spin theory. As we recall from section \ref{sec:encoding:free:conformal_scalar}, the boundary data for a conformally coupled massless spin-0 field has conformal weights $\Delta_\pm = 2,1$, which generate antipodally even and odd solutions, respectively. This relation between antipodal symmetry and boundary data, which holds only for massless fields, will be crucial for our construction.

Consider now the higher-spin dS/CFT of \cite{Anninos:2011ui}, for either the type-A or type-B model, with the choice of boundary conditions that corresponds to a free CFT. As we've seen in section \ref{sec:encoding:interacting:HS}, the sources in the CFT partition function are then given by the $dS_4$ boundary data that is not constrained to vanish under the relevant antipodal symmetry. Thus, for the even scalar in the type-A model, the CFT source is the boundary data $\varphi_{\Delta_+}$ with weight $\Delta_+ = 2$, while for the odd pseudoscalar in the type-B model, it is $\varphi_{\Delta_-}$ with weight $\Delta_- = 1$.

The free-field limit in the bulk corresponds to 2-point functions in the CFT, i.e. to the Gaussian piece of the CFT partition function. This Gaussian is of course easy to write down explicitly. However, our goal here is to show how, \emph{given} the CFT partition function, we can read off from it the operator algebra and Hamiltonian \eqref{eq:c_algebra}-\eqref{eq:c_Hamiltonian} seen by a $dS_4/\bbZ_2$ observer. Recalling that the algebra and Hamiltonian are encoded in the oscillator normalization coefficients $\left|N_l(\omega)\right|^2$, our task is to read off these coefficients from the CFT partition function. 

We begin by stating the result. Using the mode expansion \eqref{eq:boundary_modes} of the boundary source field and the observer's $\bbR\times O(3)$ symmetry, the Gaussian piece of the CFT partition function must have the form:
\begin{align}
 Z_\text{CFT}\big[c_{lm}(\omega),c^*_{lm}(\omega)\big] = \exp\left(-\int_0^\infty\frac{d\omega}{2\pi}\sum_{lm} S_l(\omega) \left|c_{lm}(\omega)\right|^2 \right) \ , \label{eq:Z}
\end{align}
with some coefficients $S_l(\omega)$. We now claim that the normalization coefficients $\left|N_l(\omega)\right|^2$, and with them the observer's operator algebra and Hamiltonian, can be extracted from the coefficients $S_l(\omega)$ in $Z_\text{CFT}$ as:
\begin{align}
 \left|N_l(\omega)\right|^2 = S_l(\omega)\cdot \frac{1}{2}\left(\coth\frac{\pi\omega}{2}\right)^{\eta\,(-1)^l} \ , \label{eq:S_to_N}
\end{align}
with $\eta = \pm 1$ for the type-A and type-B model, respectively. This result was derived in \cite{Hackl:2014txa} for $\eta=+1$ by directly evaluating both sides using bulk methods. For our present purposes, that sort of derivation, though always possible for free fields, is beside the point. Instead, we will now show how eq. \eqref{eq:S_to_N} can be derived \emph{holographically}, using only the dS/CFT dictionary of \cite{Maldacena:2002vr,Anninos:2011ui} and the relation between antipodal symmetry and boundary data.

We begin by working not in $dS_4/\bbZ_2$, but in its $dS_4$ double cover. We use eq. \eqref{eq:HH} to interpret $Z_\text{CFT}$ as the Hartle-Hawking wavefunctional $\Psi_\text{H.H.}$ of the bulk theory on $dS_4$, evaluated in the $\varphi_{\Delta_+}$ or $\varphi_{\Delta_-}$ basis on $\scri^+$ (in the type-A and type-B model, respectively). In the free-field limit, the bulk theory is non-gravitating, so the Hartle-Hawking wavefunctional is just the Bunch-Davies vacuum $\Psi_0$. Decomposing the $\varphi_{\Delta_+}$ or $\varphi_{\Delta_-}$ boundary data on $\scri^+$ into $(\omega,l,m)$ modes as in \eqref{eq:boundary_modes}, we get an expression of the form \eqref{eq:Z} for $Z_\text{CFT}$, a.k.a. the Bunch-Davies wavefunctional:
\begin{align}
 \Psi_0\big[c_{lm}(\omega),c^*_{lm}(\omega)\big] = Z_\text{CFT}\big[c_{lm}(\omega),c^*_{lm}(\omega)\big] = 
  \exp\left(-\int_0^\infty\frac{d\omega}{2\pi}\sum_{lm} S_l(\omega) \left|c_{lm}(\omega)\right|^2 \right) \ . \label{eq:BD_func_c}
\end{align}
Let us now introduce the oscillators $a_{lm}(\omega)$ in the observer's causal patch. While these were originally defined in $dS_4/\bbZ_2$, we can straightforwardly reinterpret them  as a basis of \emph{antipodally symmetric} modes in $dS_4$. We now invoke the relation between antipodal symmetry and boundary data for massless fields, i.e. that the basis of boundary data $\varphi_{\Delta_+}$ ($\varphi_{\Delta_-}$) is also the basis of antipodally even (odd) bulk solutions in $dS_4$. In the language of $(\omega,l,m)$ modes, this implies that the simple proportionality \eqref{eq:a_c} between $c_{lm}(\omega)$ and $a_{lm}(\omega)$ can be lifted from $dS_4/\bbZ_2$ into the $dS_4$ double cover. Note that this is only true for massless fields, since otherwise the antipodally symmetric modes $a_{lm}(\omega)$ would get a contribution from the boundary data of the other conformal weight. Using the proportionality \eqref{eq:a_c}, we can now rewrite the functional \eqref{eq:BD_func_c} in the $a_{lm}(\omega),a_{lm}^*(\omega)$ basis:
\begin{align}
 \Psi_0\big[a_{lm}(\omega),a^*_{lm}(\omega)\big] = 
  \exp\left(-\int_0^\infty\frac{d\omega}{2\pi}\sum_{lm} \frac{S_l(\omega)}{\left|N_l(\omega)\right|^2} \left|a_{lm}(\omega)\right|^2 \right) \ . \label{eq:BD_func_a}
\end{align}
Let us now apply the map \eqref{eq:state_operator} that turns wavefunctionals on $dS_4$ into operators on the causal patch. In an antipodally symmetric basis such as $a_{lm}(\omega),a_{lm}^*(\omega)$, this map takes the form of a Wigner-Weyl transform, we we've seen in eq. \eqref{eq:state_operator_antipodal} or \eqref{eq:state_operator_null}. The Wigner-Weyl transform of an operator such as \eqref{eq:BD_func_a}, i.e. a Gaussian in harmonic-oscillator modes $a,a^*$, is well-known. Up to normalization, it reads:
\begin{align}
 \exp\left(-\beta\hat a^\dagger\hat a \right) \ \longleftrightarrow \ \exp\left(-2\tanh\left(\frac{\beta}{2}\right) a^* a \right) \ , \label{eq:Gaussian_WW}
\end{align}
for any coefficient $\beta$. Thus, the causal-patch operator corresponding to the wavefunctional \eqref{eq:BD_func_a} reads:
\begin{align}
 \hat\Psi_0 = \exp\left(-2\int_0^\infty\frac{d\omega}{2\pi}\sum_{lm} \tanh^{-1}\left(\frac{S_l(\omega)}{2\left|N_l(\omega)\right|^2}\right) 
   \hat a^\dagger_{lm}(\omega)\, \hat a_{lm}(\omega)\right) \ ,
 \label{eq:BD_operator_a}
\end{align}
where, by abuse of notation, $a_{lm}(\omega)$ are again modes in the causal patch, as opposed to their global antipodally symmetric versions.

On the other hand, we know that under the map \eqref{eq:state_operator}, the Bunch-Davies wavefunctional becomes the operator \eqref{eq:Psi_BD}. In terms of the harmonic oscillators $\hat a_{lm}(\omega)$ in the causal patch, this can be written as:
\begin{align}
 \hat\Psi_0 = \hat R\, e^{-\pi\hat H} 
   = \exp\left(\int_0^\infty\frac{d\omega}{2\pi}\sum_{lm} \left(-\pi\omega + i(\arg\eta + \pi l)\right) \hat a^\dagger_{lm}(\omega)\, \hat a_{lm}(\omega)\right) \ , \label{eq:BD_operator_a_direct}
\end{align}
where the $-\pi\omega$ piece generates the $e^{-\pi\hat H}$ factor, while the $i(\arg\eta + \pi l)$ piece generates the antipodal map $\hat R$ on the 2-sphere for a field with parity $\eta=\pm 1$.

Comparing the expressions \eqref{eq:BD_operator_a} and \eqref{eq:BD_operator_a_direct}, we arrive at the result \eqref{eq:S_to_N}, using the following identity for the hyperbolic tangent of a complex argument:
\begin{align}
 \tanh\left(x + \frac{k\pi i}{2}\right) = \left(\tanh x\right)^{(-1)^k} \ .
\end{align}
To reiterate, our result \eqref{eq:S_to_N} expresses the operator algebra and Hamiltonian for an observer in $dS_4/\bbZ_2$ in terms of field modes on $\scri$. In the above derivation, we used only the CFT partition function and the relation between boundary data and antipodal symmetry, with no reference to bulk dynamics. 

\subsection{The CFT does not encode the observer's S-matrix} \label{sec:dS/CFT:S_matrix}

So far, we've managed to express holographically the observer's operator algebra and Hamiltonian. Another object of interest is the observer's \emph{S-matrix}, i.e. the transition amplitudes between her past horizon $H_i$ and her future horizon $H_f$. Note that this S-matrix is only observable \emph{asymptotically}; indeed, the observer can affect $H_i$ only in her infinite past, and can measure $H_f$ only in her infinite future. Be that as it may, we can now ask two questions:
\begin{enumerate}
 \item How is the S-matrix expressed in the language of the present section?
 \item Can it be recovered holographically from the CFT, as with the operator algebra and Hamiltonian?
\end{enumerate}
As we will show below, the answer to the first question is that the S-matrix is encoded in the complex phases of the normalization coefficients $N_l(\omega)$ from \eqref{eq:a_c}. On the other hand, the entire information in the CFT (at the 2-point level, i.e. for free bulk fields) is encapsulated in the coefficients $S_l(\omega)$ from \eqref{eq:Z}, which encode only the \emph{absolute value} of $N_l(\omega)$, as we've seen in \eqref{eq:S_to_N}. Thus, the answer to our second question appears to be ``no''. With some creative license, one might say that because $Z_\text{CFT}$ is real, it has ``no room'' to encode the S-matrix. This is reminiscent of the argument in \cite{Balasubramanian:2002zh} that the real $Z_\text{CFT}$ cannot encode complex transition amplitudes. Admittedly, for free bulk fields, the whole question of what can be ``recovered from the CFT'' is interpretation-laden, since one can always calculate everything directly in the bulk. Nevertheless, the above arguments lead us to expect that in $dS_4/\bbZ_2$ holography for the full interacting bulk theory, the notion of an observer's S-matrix will not survive.

Let us now justify the statement that the S-matrix is encoded in the phase of $N_l(\omega)$. So far, the phases of the oscillators $a_{lm}(\omega)$ in the causal patch, and thus the phases of $N_l(\omega)$, have not been fixed at all. To do that, we must write an explicit expression for $a_{lm}(\omega)$ in terms of the bulk field $\varphi(x)$ on some hypersurface. The two most natural hypersurfaces to use are the horizons $H_i,H_f$. These are the boundaries of the causal patch in $dS_4/\bbZ_2$, which get mapped to half-horizons in the $dS_4$ double cover (with the other halves bounding the antipodal causal patch). With the choice \eqref{eq:p_if} for the worldline endpoints $p_i,p_f$, the loci of these half-horizons in the $\bbR^{1,4}$ embedding space read:
\begin{align}
 H_i:\ x^\mu = (-e^{-\tau}, e^{-\tau}, \vec n) \quad ; \quad H_f:\ x^\mu = (e^\tau, e^\tau, \vec n) \ . \label{eq:H_coords}
\end{align}
Here, $\vec n$ is a 3d unit vector parameterizing the 2-sphere of lightrays that make up each horizon, while $\tau$ is a null coordinate along each ray. The observer's time-translation Killing field $\xi^\mu$ acts on the horizons as the null translation $\del_\tau$. Using the coordinates \eqref{eq:H_coords}, we can explicitly define annihilation operators for modes with frequency $\omega$ on each horizon as \cite{Halpern:2015cia}:
\begin{align}
 \begin{split}
   a^{(i)}_{lm}(\omega) &= \sqrt{2\omega}\int_{-\infty}^\infty d\tau\int_{S_2} d^2n\, e^{i\omega\tau}\, Y^*_{lm}(\vec n)\, \varphi(-e^{-\tau}, e^{-\tau}, \vec n) \ ; \\
   a^{(f)}_{lm}(\omega) &= \sqrt{2\omega}\int_{-\infty}^\infty d\tau\int_{S_2} d^2n\, e^{i\omega\tau}\, Y^*_{lm}(\vec n)\, \varphi(e^\tau, e^\tau, \vec n) \ ,
 \end{split} \label{eq:a_if}
\end{align}
where $d^2n$ is the area element on the unit 2-sphere. These two sets of annihilation operators translate into two choices of the normalization coefficients $N_l(\omega)$ in \eqref{eq:a_c}:
\begin{align}
 a^{(i)}_{lm}(\omega) = N^{(i)}_l(\omega)\,c_{lm}(\omega) \quad ; \quad a^{(f)}_{lm}(\omega) = N^{(f)}_l(\omega)\,c_{lm}(\omega) \ . \label{eq:a_c_if}
\end{align}

Now, by unitarity and $\bbR\times O(3)$ symmetry, an $(\omega,l,m)$ mode on $H_i$ must evolve into the same mode on $H_f$, up to a phase factor $\exp[i\alpha_l(\omega)]$. These phase factors will serve as our definition of the observer's S-matrix. In the free theory, they can be identified with the ratios $a^{(f)}_{lm}(\omega)/a^{(i)}_{lm}(\omega)$ in a classical solution, or, equivalently, with the ratio of the $N_l(\omega)$ factors on the two horizons:
\begin{align}
 e^{i\alpha_l(\omega)} = \frac{a^{(f)}_{lm}(\omega)}{a^{(i)}_{lm}(\omega)} = \frac{N^{(f)}_l(\omega)}{N^{(i)}_l(\omega)} \ . \label{eq:S_matrix}
\end{align}

Our next observation is that $N^{(i)}_l(\omega)$ and $N^{(f)}_l(\omega)$ are in fact related through complex conjugation. To see this, note that the boundary mode $c_{lm}(\omega)$ from \eqref{eq:boundary_modes} can be expressed using asymptotic fields on either $\scri^+$ or $\scri^-$ in the $dS_4$ double cover:
\begin{align}
 \begin{split}
   c_{lm}(\omega) &= \int_{-\infty}^\infty dt\int_{S_2} d^2n\, e^{i\omega t}\, Y^*_{lm}(\vec n)\, \varphi_\Delta(\cosh t, \sinh t, \vec n) \\
    &= \eta\,(-1)^l\int_{-\infty}^\infty dt\int_{S_2} d^2n\, e^{i\omega t}\, Y^*_{lm}(\vec n)\, \varphi_\Delta(-\cosh t, -\sinh t, \vec n) \ ,
 \end{split} \label{eq:c_pm}
\end{align}
where we used the antipodal symmetry of the boundary field $\varphi_\Delta(-\ell) = \eta \varphi_\Delta(\ell)$ and of spherical harmonics $Y_{lm}(-\vec n) = (-1)^l\,Y_{lm}(\vec n)$. Now, we can read off from \eqref{eq:a_if},\eqref{eq:c_pm} that time reflection $x^0\rightarrow -x^0$ in the $\bbR^{1,4}$ embedding space acts on the boundary and horizon modes as follows:
\begin{align}
 c_{lm}(\omega) \rightarrow \eta\,(-1)^l\, c^*_{l,-m}(\omega) \ ; \quad a^{(i)}_{lm}(\omega) \rightarrow a^{(f)*}_{l,-m}(\omega) \ ; \quad 
 a^{(f)}_{lm}(\omega) \rightarrow a^{(i)*}_{l,-m}(\omega)
\end{align}
Since the boundary-to-bulk propagation of our antipodally symmetric free field $\varphi$ is invariant under $x^0\rightarrow -x^0$, we conclude the following relation between the $N_l(\omega)$ normalization coefficients:
\begin{align}
 N_l^{(i)}(\omega) = \eta\,(-1)^l\,N_l^{(f)*}(\omega) \ .
\end{align}
Plugging this into \eqref{eq:S_matrix}, we conclude that the phase factors comprising the S-matrix are indeed encoded in the complex phase of $N_l(\omega)$. For instance, in terms of the $N_l(\omega)$'s on the future horizon, the S-matrix phases \eqref{eq:S_matrix} read:
\begin{align}
 \alpha_l(\omega) = \arg\eta + \pi l + 2\arg N^{(f)}_l(\omega) \ .
\end{align}
As discussed above, these phases are \emph{not} captured by our dS/CFT construction. Once again, in the free bulk theory, this does not prevent us from finding them explicitly. This calculation was performed in \cite{Hackl:2014txa} for the antipodally even ($\eta = +1$) conformally-coupled massless scalar. In that case, the full complex normalization coefficients between the boundary modes and the future horizon modes read:
\begin{align}
 N^{(f)}_l(\omega) = -\frac{i}{\sqrt{2\omega}}\prod_{k=0}^l \left(\omega + i(-1)^{l+k+1}k \right)^{(-1)^{l+k+1}} \ .
\end{align}

\section{Discussion} \label{sec:discuss}

In this paper, we argued for an approach to dS/CFT that combines the higher-spin model of \cite{Anninos:2011ui} with the idea of identifying antipodal points in the bulk. We made progress towards this goal on several fronts. First, we studied the relationship between antipodal symmetry and boundary data for classical theories on $dS_4$. This relationship becomes especially simple for free massless fields and for interacting higher-spin theory, where boundary data with a particular conformal weight corresponds to a particular antipodal symmetry sign. Second, we developed and motivated a framework for (non-gravitational) quantum field theory in antipodally-identified de Sitter space $dS_4/\bbZ_2$; we saw that a standard quantum description does not exist globally, but exists for every observer inside her causal patch; we also saw that observers may either agree or disagree on expectation values for the ``same'' observables, depending on whether they agree on the arrow of time in the relevant spacetime regions. Finally, for free massless bulk fields, i.e. the linearized limit of higher-spin theory, we constructed a holographic description of an observer's operator algebra and Hamiltonian in terms of the boundary CFT; on the other hand, our construction apparently fails to reproduce the observer's S-matrix, since the CFT partition function does not contain the necessary complex phases.

The main goal for future work is to extend the holographic construction of section \ref{sec:dS/CFT} from free massless fields to interacting higher-spin gravity. To appreciate what this might involve, let us recall which special properties of free massless fields came into play. First, we used the relationship between antipodal symmetry and boundary data, which appears to extend also to interacting higher-spin theory. Second, we used the linearity of the field equations in order to ``propagate'' the observer's quantum world-picture from the causal patch onto the boundary. As we've seen, it is not clear how to do this in an interacting theory. Our hope is that higher-spin theory will prove easier in this context than arbitrary interacting theories. Indeed, recall that our problem is that the boundary $\scri$ intersects the causal patch at only two points $p_i,p_f$. Now, it so happens that the higher-spin gravity field equations are given in an ``unfolded'' formulation, where the entire solution is encoded via master fields \emph{at any single spacetime point}. This should make it possible to do holography for each observer using only the endpoints $p_i,p_f$ of her causal patch, instead of the entire boundary. The boundary itself will then serve as an underlying global framework for comparing the world-pictures of different observers. In this sense, we expect that $\scri$ will take over the role played in section \ref{sec:QFT} by the $dS_4$ double cover of the bulk spacetime.

The next difficulty in upgrading to full-fledged higher-spin theory is that we've been heavily using the specific causal structure of $dS_4$ and $dS_4/\bbZ_2$, in particular the geometry of horizons. In General Relativity, the causal structure becomes dynamical. In higher-spin theory, things appear at first to be even worse: the metric, and with it the lightcone structure, is demoted to a gauge-dependent component of the higher-spin gauge connection. This, along with the apparently non-local nature of the interactions, calls into question the whole notion of a causal structure in the theory. That is potentially a severe problem for our entire motivation. Indeed, our primary goal was to understand quantum gravity inside horizons; this led us to study de Sitter space, in which the best candidate for some semblance of quantum gravity is higher-spin theory. But what is it all good for, if higher-spin theory ends up not knowing what a horizon is?

A possible way out lies in the recent formulation \cite{Neiman:2015wma} of higher-spin theory by one of the authors, which allows it to be viewed as a gauge theory on a fixed (anti) de Sitter background. In this formulation, pure $dS_4$ with its metric and causal structure is retained, even though we are still in interacting, non-perturbative, higher-spin theory. Thus, higher-spin theory is potentially simpler than General Relativity: not only can a causal structure be defined, but it can be made non-dynamical! This also suggests that higher-spin theory evades a crucial flaw in the idea of antipodal identification, as applied to General Relativity. In GR, fluctuations over pure $dS_4/\bbZ_2$ necessarily create closed timelike curves \cite{Gao:2000ga,Leblond:2002ns,Parikh:2002py}. This doesn't happen for higher-spin theory, if we can always view it as living on a fixed $dS_4$ or $dS_4/\bbZ_2$ geometry. The crucial open question is whether the $dS_4$ or $dS_4/\bbZ_2$ horizons in the higher-spin theory of \cite{Neiman:2015wma} in fact \emph{behave} as causal boundaries for field propagation. Addressing this issue would require a careful understanding of causality in unfolded dynamics. 

Another open problem is to develop the necessary language for asymptotics and holography within the fixed-background formulation \cite{Neiman:2015wma} of higher-spin theory. As an added benefit, this may provide new insight on the AdS$_4$/CFT$_3$ duality between higher-spin gravity and free vector models. To understand the link between these issues, note that the higher-spin/free-CFT duality hinges on the existence of boundary conditions that preserve higher-spin symmetry to all orders in the interaction \cite{Vasiliev:2012vf}. In Vasiliev's argument \cite{Vasiliev:2012vf} that such boundary conditions exist, he made use of a parity operator that involves a reflection $z\rightarrow -z$ in Poincare coordinates. In the $AdS_4$ context of \cite{Vasiliev:2012vf}, this was termed a ``doubling'' of the bulk spacetime. However, in $dS_4$, one can see from eq. \eqref{eq:x} that this coordinate transformation is just the antipodal map (and is of CT type, as opposed to P type in $AdS_4$). Thus, the higher-spin/free-CFT duality in $AdS_4$ can be linked to antipodal symmetry in $dS_4$. The fixed-background formulation of higher-spin theory \cite{Neiman:2015wma} seems especially well-suited for the study of antipodal symmetry, since it has an invariant notion of antipodal points as part of the fixed $dS_4$ geometry.

Finally, we stress that quantization of higher-spin theory is not yet understood even perturbatively, except through AdS/CFT. Our hope is essentially that the kind of holography discussed in this paper will provide a new method for quantizing the theory, which would be applicable to de Sitter causal patches.

\section*{Acknowledgements}

We are grateful to Laurent Freidel, Rob Myers and Vasudev Shyam for discussions. During early stages of this work, IFH was based at Perimeter Institute. Research at Perimeter Institute is supported by the Government of Canada through Industry Canada and by the Province of Ontario through the Ministry of Research \& Innovation. YN also acknowledges support of funding from NSERC Discovery grants. 

\appendix
\section{Proof of Theorems \ref{thm:delta_asymptotics} and \ref{thm:gauge_asymptotics}} \label{app:delta_asymptotics}

We will use the Poincare coordinates \eqref{eq:x},\eqref{eq:l} for the bulk point $x$ and the boundary point $l\in\scri^+$ to which it asymptotes. Without loss of generality, we can place $x$ and $l$ at the origin of $\vec r$ space:
\begin{align}
 x^\mu = \frac{1}{2z}\left(1 - z^2, -1 - z^2, \vec 0 \right) \quad ; \quad l^\mu = \frac{1}{2}\left(1,-1,\vec 0 \right) \ , \label{eq:x_origin}
\end{align}
while keeping the boundary source point $\ell\in\scri^+$ at general $\vec r$:
\begin{align}
 \ell^\mu = \left(\frac{r^2 + 1}{2}, \frac{r^2 - 1}{2}, \vec r \right) \ . \label{eq:ell}
\end{align}
With these choices, the inner product $x\cdot\ell$ reads:
\begin{align}
 x\cdot\ell = \frac{z^2 - r^2}{2z} \ . \label{eq:x_dot_ell}
\end{align}

\subsection{Asymptotics of $\delta^{(k)}(x\cdot\ell)$}

Consider the delta-function derivative $\delta^{(k)}(x\cdot\ell)$ with $k\geq 0$. To evaluate it as a distribution on $\scri^+$, we integrate the boundary source point \eqref{eq:ell} against a test function $f(\ell)=f(\vec r)$, using the flat volume measure $d^3\ell=d^3r$ on the lightcone section \eqref{eq:ell}:
\begin{align}
 \int d^3\ell\, \delta^{(k)}(x\cdot\ell) f(\ell) = \int d^3r\, \delta^{(k)}\!\left(\frac{z^2 - r^2}{2z}\right) f(\vec r) 
   = (2z)^{k+1}\int d^3r\, \delta^{(k)}\!\left(z^2 - r^2\right) f(\vec r) \ .
\end{align}
Switching variables to $u \equiv r^2$ and $\vec n \equiv \vec r/r$, this becomes:
\begin{align}
 \begin{split}
   \int d^3\ell\, \delta^{(k)}(x\cdot\ell) f(\ell) &= 2^k z^{k+1}\,\int_0^\infty du\sqrt{u}\, \delta^{(k)}\!\left(z^2 - u\right)\int_{S_2} d^2n\, f\big(\sqrt{u}\,\vec n\big) \\
     &= 2^k z^{k+1}\int_{S_2} d^2n \left.\frac{d^k}{du^k}\left(\sqrt{u}\,f\big(\sqrt{u}\,\vec n\big) \right) \right|_{u = z^2} \ . \label{eq:u_integral}
 \end{split}
\end{align}
where $d^2n$ is the area element on the unit 2-sphere. To analyze the expression \eqref{eq:u_integral}, consider the Taylor expansion of the test function $f(\vec r)$. From the zeroth-order term, we get:
\begin{align}
 4\pi\cdot 2^k z^{k+1} f(\vec 0) \left.\frac{d^k}{du^k}\left(\sqrt{u}\right)\right|_{u = z^2} = 4\pi z^{2-k} f(\vec 0) \times \left\{
  \begin{array}{ll}
    1 & \quad k = 0 \\
    (-1)^{k-1}(2k-3)!! & \quad k \geq 1
  \end{array} \right. \ .
\end{align}
The $N^{\text{th}}$-order term in the Taylor expansion contributes an extra factor of $u^{N/2}$ to the integrand, which translates into a factor of $z^N$ in the result. Only even $N$ contribute in this way, since the $d^2n$ integral on the 2-sphere vanishes for odd $N$. This concludes the proof of eqs. \eqref{eq:delta_asymptotics}-\eqref{eq:delta_k_asymptotics} in Theorem \ref{thm:delta_asymptotics}.

\subsection{Asymptotics of $\theta(x\cdot\ell)$}

We proceed similarly, integrating the boundary source point \eqref{eq:ell} against a test function $f(\ell)=f(\vec r)$:
\begin{align}
 \int d^3\ell\, \theta(x\cdot\ell) f(\ell) = \int d^3r\, \theta\!\left(\frac{z^2 - r^2}{2z}\right) f(\vec r)
  = \int d^3r\, \theta\!\left(z^2 - r^2\right) f(\vec r) \ .
\end{align}
Using spherical coordinates $r$ and $\vec n = \vec r/r$, this becomes:
\begin{align}
 \int d^3\ell\, \theta(x\cdot\ell) f(\ell) = \int_0^z r^2 dr \int_{S_2} d^2n\, f(r\vec n) \ .
\end{align}
The zeroth-order term in the Taylor expansion of $f(\vec r)$ gives:
\begin{align}
 4\pi f(\vec 0) \int_0^z r^2 dr = \frac{4\pi}{3}\,z^3 f(\vec 0) \ .
\end{align}
As before, the $N^{\text{th}}$-order term in the Taylor expansion contributes an extra factor of $r^N$ to the integrand, which translates into a factor of $z^N$ in the result. Only even $N$ contribute, since the odd-$N$ terms vanish after the $d^2n$ spherical integral. This concludes the proof of eq. \eqref{eq:theta_asymptotics} in Theorem \ref{thm:delta_asymptotics}.

\subsection{Proof of Theorem \ref{thm:gauge_asymptotics}}

We now consider the (partially) massless gauge field propagator \eqref{eq:phi_lambda}-\eqref{eq:G}. We again parametrize the boundary point $\ell$ as in \eqref{eq:ell}, fixing $x$ and $l$ as in \eqref{eq:x_origin}. The polarization vector $\lambda^\mu(\ell)=\lambda^\mu(\vec r)$ can then be parameterized by a null complex 3d vector $\vec\lambda(\vec r)$, as:
\begin{align}
 \lambda^\mu = (\vec r\cdot\vec\lambda,\ \vec r\cdot\vec\lambda,\ \vec\lambda) \ .
\end{align}
We wish to evaluate the smearing integral:
\begin{align}
 \begin{split}
   &\int d^3\ell\, G^{\mu_1\dots\mu_n}_{\nu_1\dots\nu_n}(x;\ell)\,\lambda^{\nu_1}(\vec r)\dots\lambda^{\nu_s}(\vec r) \\
   &\quad = \int d^3r \left((x\cdot\lambda)\ell^{\mu_1} - (x\cdot\ell)\lambda^{\mu_1}\right)\dots ((x\cdot\lambda)\ell^{\mu_s} - (x\cdot\ell)\lambda^{\mu_s})\,\delta^{(s+j)}(x\cdot\ell) \ . 
 \end{split} \label{eq:G_integral}
\end{align}
Through the same mechanisms as above, the leading contribution at small $z$ comes from picking out the lowest power of $\vec r$ everywhere except in $x\cdot\ell$. Higher powers of $\vec r$ will add positive even powers of $z$. Also, since we're interested in the result only up to $l^\mu$-parallel terms, we can subtract $l^\mu$ from all the $\ell^\mu$'s with free indices (which would otherwise have been the piece of $\ell^\mu$ with the lowest power of $\vec r$). Thus, to find the leading-order result, we replace the factors of $\ell^\mu$, $\lambda^\mu$ and $x\cdot\lambda$ in \eqref{eq:G_integral} with:
\begin{align}
 \ell^\mu \rightarrow \left(0,0,\vec r\right) \equiv r^\mu \ ; \quad \lambda^\mu \rightarrow \left(0,0,\vec\lambda(\vec 0)\right) \ ; \quad 
 x\cdot\lambda \rightarrow -\frac{\vec r\cdot\vec\lambda(\vec 0)}{z} \ ,
\end{align}
while keeping the exact expression \eqref{eq:x_dot_ell} for $x\cdot\ell$. From now on, we omit the argument in $\vec\lambda(\vec 0)$, with the understanding that $\vec\lambda$ is evaluated at the origin of $\vec r$ space. 

Let us now expand the binomial in \eqref{eq:G_integral}, focusing on a term with $k$ factors of $(x\cdot\lambda)\ell^{\mu}$ (e.g. the \emph{first} $k$ factors) and $s-k$ factors of $(x\cdot\ell)\lambda^{\mu}$. With the above remarks all taken into account, this term becomes:
\begin{align}
 \begin{split}
   &(-1)^{s-k}\int d^3r \, \delta^{(s+j)}(x\cdot\ell)\, (x\cdot\lambda)^k (x\cdot\ell)^{s-k}\, r^{\mu_1}\dots r^{\mu_k}\lambda^{\mu_{k+1}}\dots\lambda^{\mu_s} \\
    &\quad = \frac{2^{j+k+1} (-1)^k (s+j)!}{(j+k)!}\,z^{j+1} \int d^3r\, \delta^{(j+k)}(z^2-r^2)\,(\vec r\cdot\vec\lambda)^k\, 
    r^{\mu_1}\dots r^{\mu_k}\lambda^{\mu_{k+1}}\dots\lambda^{\mu_s} \ . 
 \end{split}
\end{align}
Using spherical coordinates $r$ and $\vec n = \vec r/r$ with $n^\mu \equiv (0,0,\vec n)$, this becomes:
\begin{align}
 \begin{split}
   \frac{2^{j+k+1} (-1)^k (s+j)!}{(j+k)!}\,z^{j+1} &\int_0^\infty dr\,r^{2k+2}\, \delta^{(j+k)}(z^2-r^2) \\
   \times &\int_{S_2} d^2n\, (\vec n\cdot\vec\lambda)^k\, 
     n^{\mu_1}\dots n^{\mu_k}\lambda^{\mu_{k+1}}\dots\lambda^{\mu_s} \ . \label{eq:radial_and_angular}
 \end{split}
\end{align}
With the change of variables $u=r^2$, the radial integral evaluates to:
\begin{align}
 \begin{split}
   \int_0^\infty dr\,r^{2k+2}\, \delta^{(j+k)}(z^2-r^2) &= \frac{1}{2}\int_0^\infty du\,u^{k+1/2}\,\delta^{(j+k)}(z^2-u) 
    = \frac{1}{2}\left.\frac{d^{j+k}}{du^{j+k}}\left(u^{k+1/2}\right)\right|_{u = z^2} \\
    &= \frac{(-1)^{j-1}(2j-3)!!(2k+1)!!}{2^{j+k+1}}\,z^{1-2j} \ .
 \end{split}
\end{align}
As for the angular integral in \eqref{eq:radial_and_angular}, it evaluates to:
\begin{align}
 \int_{S_2} d^2n\, (\vec n\cdot\vec\lambda)^k\, n^{\mu_1}\dots n^{\mu_k} = \frac{4\pi k!}{(2k+1)!!}\,\lambda^{\mu_1}\dots\lambda^{\mu_k} \ . \label{eq:angular}
\end{align}
This can be shown in three steps. First, we note the following trace identity in 3 dimensions:
\begin{align}
 \delta_{i_1 j_1}\dots\delta_{i_k j_k}\delta^{(i_1 j_1}\dots\delta^{i_k j_k)} = \frac{1}{2k+1} \ ,
\end{align}
which is easy to prove recursively in $k$. Then, by using rotational symmetry and tracing both sides with $\delta_{i_1 j_1}\dots\delta_{i_k j_k}$, it is straightforward to evaluate:
\begin{align}
 \int_{S_2} d^2n\, n^{i_1}\dots n^{i_k} n^{j_1}\dots n^{j_k} = \frac{4\pi}{2k+1}\,\delta^{(i_1 j_1}\dots\delta^{i_k j_k)} \ .
\end{align}
Now, contracting with $k$ factors of $\vec\lambda$, and discarding all terms in which the null $\vec\lambda$ gets contracted with itself, we obtain the result \eqref{eq:angular}.

Overall, the expression \eqref{eq:radial_and_angular} evaluates to:
\begin{align}
 \frac{4\pi (-1)^{j+k+1}k!(s+j)!(2j-3)!!}{(j+k)!}\,z^{2-j}\,\lambda^{\mu_1}\dots\lambda^{\mu_s} \ .
\end{align}
Summing over $k$ with the binomial coefficient $\binom{s}{k}$, we obtain the result of Theorem \ref{thm:gauge_asymptotics}.

\section{From global states on a horizon to operators on a half-horizon} \label{app:horizon_map} 
 
In this Appendix, we construct the map \eqref{eq:state_operator} between Hilbert-space vectors on $dS_4$ and operators on a causal patch $D$, in terms of an antipodally symmetric basis on one of the horizons $H_i,H_f$. In other words, we construct the analog of eq. \eqref{eq:state_operator_antipodal}, with a null horizon in place of the spatial slice $\Sigma$. To avoid a clutter of plus/minus signs, we focus on the future horizon $H_f$. Our construction is an adaptation of results from \cite{Halpern:2015cia}. We ignore the infamous issues with zero modes on null hypersurfaces \cite{Maskawa:1975ky,Yamawaki:1997cj}.

Let us choose a frame in the $\bbR^{1,4}$ embedding space such that the asymptotic endpoints of the causal patch read:
\begin{align}
 p_i^\mu = (-1,1,\vec 0) \quad ; \quad p_f^\mu = (1,1,\vec 0) \ .
\end{align}
We can then coordinatize the horizon $H_f$ as:
\begin{align}
 x^\mu = (u,u,\vec n) \ . \label{eq:horizon_coords}
\end{align}
The horizon consists of lightrays labeled by the 3d unit vector $\vec n$, with $u$ an affine null coordinate along each ray. The bifurcation surface $H_i\cap H_f$ is at $u=0$. The full horizon with $-\infty<u<\infty$ causally spans all of $dS_4$, while the half-horizons $u\gtrless0$ span the causal patch $D$ and its antipode $\overbar D$, respectively. The antipodal map sends the horizon $H_f$ to itself, with individual points transforming as:
\begin{align}
 \calA: \ (u,\vec n) \rightarrow (-u,-\vec n) \ .
\end{align}

The phase space on $H_f$ is spanned by the values $\varphi(x)$ of the field on it. The field's normal derivative doesn't need to be specified separately, since the normal to a null horizon is also tangent to it. The commutator for the fields on $H_f$ reads:
\begin{align}
 \left[\hat \varphi(u,\vec n), \hat \varphi(u',\vec n')\right] = \frac{i}{4}\,\delta^{(2)}(\vec n,\vec n')\sign(u'-u) \ , \label{eq:horizon_commutator}
\end{align}
where $\delta^{(2)}(\vec n,\vec n')$ is a delta function on the unit 2-sphere of lightrays. We see that fields on the same lightray do not commute. There are some caveats concerning the applicability of the commutator \eqref{eq:horizon_commutator} in interacting theories. Those are summarized in \cite{Halpern:2015cia}, and we will ignore them here.

To write down Hilbert-space vectors as wavefunctionals, we require a maximal commuting set of field operators. Such a set is given by the antipodally symmetric component of $\varphi(x)$ on the horizon:
\begin{align}
 \varphi_\text{sym}(u,\vec n) = \frac{\varphi(u,\vec n) + \eta \varphi(-u,-\vec n)}{2} \ , \label{eq:phi_sym_H}
\end{align}
where $\eta$ is the field's intrinsic parity as before. For future use, we also introduce a notation for the field component with the opposite antipodal symmetry:
\begin{align}
 \varphi_\text{skew}(u,\vec n) = \frac{\varphi(u,\vec n) - \eta \varphi(-u,-\vec n)}{2} \ , \label{eq:phi_skew_H}
\end{align}

The antipodally symmetric field $\varphi_\text{sym}(u,\vec n)$ on the entire horizon $H_f$ can be parameterized by its values on the half-horizon $u>0$ spanning the causal patch. Thus, wavefunctionals on $H_f$ can be defined as $\Psi[\varphi_\text{sym}(u,\vec n)]$, with $u>0$. The antipodal map in the basis \eqref{eq:phi_sym_H} reads:
\begin{align}
 \calA\Psi[\varphi_\text{sym}(x)] = \Psi^*[\varphi_\text{sym}(x)] \ . \label{eq:A_antipodal_H} 
\end{align}
Thus, antipodally symmetric states have real wavefunctionals. 

We will require a similar parametrization for the Hilbert space on the half-horizon $u>0$. First, introduce a new null coordinate $\tau=\ln u$, such that $-\infty<\tau<\infty$ spans the half-horizon. Note that $\del_\tau$ is the horizon value of the Killing vector that generates time translations in the causal patch. A maximal set of commuting operators on the half-horizon $u>0$ is given e.g. by the $\tau$-odd modes of the field $\varphi(u,\vec n) = \varphi(e^\tau,\vec n)$:
\begin{align}
 B_D(\omega,\vec n) = 2\sqrt{\omega}\int_{-\infty}^\infty d\tau \sin(\omega\tau)\, \varphi(e^\tau,\vec n) \ .
\end{align}
Similarly, a maximal set of commuting operators on the antipodal half-horizon $u<0$ is given by:
\begin{align}
 B_{\overbar D}(\omega,\vec n) = 2\eta\sqrt{\omega}\int_{-\infty}^\infty d\tau \sin(\omega\tau)\, \varphi(-e^\tau,-\vec n) \ ,
\end{align}
where we used $\tau = \ln(-u)$ as a (past-pointing) null coordinate spanning the range $u<0$. Thus, wavefunctionals on the half-horizons $u\gtrless 0$ can be defined as $\Psi_D[B_D(\omega,\vec n)]$ and $\Psi_{\overbar D}[B_{\overbar D}(\omega,\vec n)]$ respectively, where the frequency $\omega$ is in the range $\omega>0$.

The antipodally symmetrized fields \eqref{eq:phi_sym_H} on the full horizon can also be spanned by Fourier modes with respect to $\tau$:
\begin{align}
 \begin{split}
   A_\text{sym}(\omega,\vec n) &= 2\sqrt{\omega}\int_{-\infty}^\infty d\tau \cos(\omega\tau)\, \varphi_\text{sym}(e^\tau,\vec n) \ ; \\
   B_\text{sym}(\omega,\vec n) &= 2\sqrt{\omega}\int_{-\infty}^\infty d\tau \sin(\omega\tau)\, \varphi_\text{sym}(e^\tau,\vec n) 
     = \frac{B_D(\omega,\vec n) + B_{\overbar D}(\omega,\vec n)}{2} \ ,
 \end{split} \label{eq:A_B_sym}
\end{align}
and likewise for the fields \eqref{eq:phi_skew_H} with the opposite antipodal symmetry:
\begin{align}
 \begin{split}
   A_\text{skew}(\omega,\vec n) &= 2\sqrt{\omega}\int_{-\infty}^\infty d\tau \cos(\omega\tau)\, \varphi_\text{skew}(e^\tau,\vec n) \ ; \\
   B_\text{skew}(\omega,\vec n) &= 2\sqrt{\omega}\int_{-\infty}^\infty d\tau \sin(\omega\tau)\, \varphi_\text{skew}(e^\tau,\vec n)
     = \frac{B_D(\omega,\vec n) - B_{\overbar D}(\omega,\vec n)}{2} \ ,
 \end{split} \label{eq:A_B_skew} 
\end{align}
One can verify using \eqref{eq:horizon_commutator} that the non-vanishing commutators among the global modes \eqref{eq:A_B_sym}-\eqref{eq:A_B_skew} read:
\begin{align}
 \left[\hat A_\text{sym}(\omega,\vec n), \hat B_\text{skew}(\omega',\vec n') \right] = \left[\hat A_\text{skew}(\omega,\vec n), \hat B_\text{sym}(\omega',\vec n') \right] 
   = \pi i\,\delta(\omega-\omega')\,\delta^{(2)}(\vec n, \vec n') \ .
 \label{eq:A_B_conjugates}
\end{align}

To sum up, we have an antipodally symmetric basis $\left|A_\text{sym},B_\text{sym}\right>$ for the global Hilbert space, as well as a basis $\left|B_D,B_{\overbar D}\right>$ that splits the degrees of freedom into those in $D$ and those in $\overbar D$. The bases are related through $B_\text{sym} = (B_D + B_{\overbar D})/2$, together with the fact that $A_\text{sym}$ and $B_D - B_{\overbar D} = 2B_\text{skew}$ are canonical conjugates. Plugging these observations into the prescription \eqref{eq:state_operator}, we get the following map between global states in the antipodally symmetric basis and operators on the Hilbert space of $D$:
\begin{align}
 \big\langle A_\text{sym}, B_\text{sym} \big|\Psi\big\rangle
  = \int\calD B_\text{skew}\,e^{2iB_\text{skew}\cdot A_\text{sym}}\,\big\langle B_\text{sym} + B_\text{skew} \big| \hat\Psi \big| B_\text{sym} - B_\text{skew} \big\rangle \ .
 \label{eq:state_operator_null}
\end{align}
Here, the matrix element on the RHS is evaluated between two $B_D$ mode configurations $B_\text{sym} \pm B_\text{skew}$ on the $u>0$ half-horizon. The scalar product $B_\text{skew}\cdot A_\text{sym}$ stands for:
\begin{align}
 B_\text{skew}\cdot A_\text{sym} \equiv \int_{S_2} d^2n \int_0^\infty \frac{d\omega}{2\pi}\,B_\text{skew}(\omega,\vec n) A_\text{sym}(\omega,\vec n) \ ,
\end{align}
where $d^2n$ is the area element on the unit 2-sphere.

Like its spatial-slice counterpart \eqref{eq:state_operator_antipodal}, the map \eqref{eq:state_operator_null} has the structure of a Wigner-Weyl transform, this time in a momentum basis, as in:
\begin{align}
 f(q,p) = \int \frac{dp'}{2\pi}\, e^{ip' q}\, \big\langle p + \frac{p'}{2} \big| \hat f \big| p - \frac{p'}{2} \big\rangle \ . \label{eq:WW_momentum}
\end{align}
The map \emph{becomes} a Wigner-Weyl transform if we replace the antipodally symmetrized field $\varphi_\text{sym}$ in \eqref{eq:A_B_sym} with the field $\varphi$ itself on the half-horizon $u>0$. With this replacement, $A_\text{sym}(\omega,\vec n)$ and $B_\text{sym}(\omega,\vec n)$ become canonical conjugates that span the phase space on the half-horizon.


\begin{thebibliography}{99}

\bibitem{Aharony:1999ti}
  O.~Aharony, S.~S.~Gubser, J.~M.~Maldacena, H.~Ooguri and Y.~Oz,
  ``Large N field theories, string theory and gravity,''
  Phys.\ Rept.\  {\bf 323}, 183 (2000)
  [arXiv:hep-th/9905111].

\bibitem{Witten:1998qj} 
  E.~Witten,
  ``Anti-de Sitter space and holography,''
  Adv.\ Theor.\ Math.\ Phys.\  {\bf 2}, 253 (1998)
  [hep-th/9802150].

\bibitem{Strominger:2001pn} 
  A.~Strominger,
  ``The dS / CFT correspondence,''
  JHEP {\bf 0110}, 034 (2001)
  [hep-th/0106113].
  
\bibitem{Maldacena:2002vr} 
  J.~M.~Maldacena,
  ``Non-Gaussian features of primordial fluctuations in single field inflationary models,''
  JHEP {\bf 0305}, 013 (2003)
  [astro-ph/0210603].
  
\bibitem{Harlow:2011ke} 
  D.~Harlow and D.~Stanford,
  ``Operator Dictionaries and Wave Functions in AdS/CFT and dS/CFT,''
  arXiv:1104.2621 [hep-th].
  
\bibitem{Hartle:1983ai} 
  J.~B.~Hartle and S.~W.~Hawking,
  ``Wave Function of the Universe,''
  Phys.\ Rev.\ D {\bf 28}, 2960 (1983).
  
\bibitem{Gibbons:1977mu} 
  G.~W.~Gibbons and S.~W.~Hawking,
  ``Cosmological Event Horizons, Thermodynamics, and Particle Creation,''
  Phys.\ Rev.\ D {\bf 15}, 2738 (1977).
  
\bibitem{Bekenstein:1980jp} 
  J.~D.~Bekenstein,
  ``A Universal Upper Bound on the Entropy to Energy Ratio for Bounded Systems,''
  Phys.\ Rev.\ D {\bf 23}, 287 (1981).
  
\bibitem{Bousso:1999xy} 
  R.~Bousso,
  ``A Covariant entropy conjecture,''
  JHEP {\bf 9907}, 004 (1999)
  [hep-th/9905177].
  
\bibitem{Anninos:2011ui} 
  D.~Anninos, T.~Hartman and A.~Strominger,
  ``Higher Spin Realization of the dS/CFT Correspondence,''
  arXiv:1108.5735 [hep-th].
  
\bibitem{Vasiliev:1995dn} 
  M.~A.~Vasiliev,
  ``Higher spin gauge theories in four-dimensions, three-dimensions, and two-dimensions,''
  Int.\ J.\ Mod.\ Phys.\ D {\bf 5}, 763 (1996)
  [hep-th/9611024].
  
\bibitem{Didenko:2014dwa} 
  V.~E.~Didenko and E.~D.~Skvortsov,
  ``Elements of Vasiliev theory,''
  arXiv:1401.2975 [hep-th].
  
\bibitem{Klebanov:2002ja} 
  I.~R.~Klebanov and A.~M.~Polyakov,
  ``AdS dual of the critical O(N) vector model,''
  Phys.\ Lett.\ B {\bf 550}, 213 (2002)
  [hep-th/0210114].

\bibitem{Sezgin:2003pt} 
  E.~Sezgin and P.~Sundell,
  ``Holography in 4D (super) higher spin theories and a test via cubic scalar couplings,''
  JHEP {\bf 0507}, 044 (2005)
  [hep-th/0305040].

\bibitem{Giombi:2012ms} 
  S.~Giombi and X.~Yin,
  ``The Higher Spin/Vector Model Duality,''
  J.\ Phys.\ A {\bf 46}, 214003 (2013)
  [arXiv:1208.4036 [hep-th]].
  
\bibitem{Folacci:1986gr} 
  A.~Folacci and N.~G.~Sanchez,
  ``Quantum Field Theory and the 'Elliptic Interpretation' of De Sitter Space-time,''
  Nucl.\ Phys.\ B {\bf 294}, 1111 (1987).
  
\bibitem{Parikh:2002py} 
  M.~K.~Parikh, I.~Savonije and E.~P.~Verlinde,
  ``Elliptic de Sitter space: dS/Z(2),''
  Phys.\ Rev.\ D {\bf 67}, 064005 (2003)
  [hep-th/0209120].

\bibitem{Parikh:2004ux} 
  M.~K.~Parikh and E.~P.~Verlinde,
  ``De sitter space with finitely many states: A Toy story,''
  hep-th/0403140.

\bibitem{Parikh:2004wh} 
  M.~K.~Parikh and E.~P.~Verlinde,
  ``De Sitter holography with a finite number of states,''
  JHEP {\bf 0501}, 054 (2005)
  [hep-th/0410227].

\bibitem{Neiman:2014npa} 
  Y.~Neiman,
  ``Antipodally symmetric gauge fields and higher-spin gravity in de Sitter space,''
  JHEP {\bf 1410}, 153 (2014)
  [arXiv:1406.3291 [hep-th]].
  
\bibitem{Higuchi:1986wu} 
  A.~Higuchi,
  ``Symmetric Tensor Spherical Harmonics on the $N$ Sphere and Their Application to the De Sitter Group SO($N$,1),''
  J.\ Math.\ Phys.\  {\bf 28}, 1553 (1987)
  [Erratum-ibid.\  {\bf 43}, 6385 (2002)].
  
\bibitem{Deser:2001us} 
  S.~Deser and A.~Waldron,
  ``Partial masslessness of higher spins in (A)dS,''
  Nucl.\ Phys.\ B {\bf 607}, 577 (2001)
  [hep-th/0103198].
    
\bibitem{Zinoviev:2001dt} 
  Y.~M.~Zinoviev,
  ``On massive high spin particles in AdS,''
  hep-th/0108192.
  
\bibitem{Hackl:2014txa} 
  L.~Hackl and Y.~Neiman,
  ``Horizon complementarity in elliptic de Sitter space,''
  Phys.\ Rev.\ D {\bf 91}, no. 4, 044016 (2015)
  [arXiv:1409.6753 [hep-th]].

\bibitem{Ashtekar:2014zfa} 
  A.~Ashtekar, B.~Bonga and A.~Kesavan,
  ``Asymptotics with a positive cosmological constant: I. Basic framework,''
  Class.\ Quant.\ Grav.\  {\bf 32}, no. 2, 025004 (2015)
  [arXiv:1409.3816 [gr-qc]].
  
\bibitem{Ng:2012xp} 
  G.~S.~Ng and A.~Strominger,
  ``State/Operator Correspondence in Higher-Spin dS/CFT,''
  Class.\ Quant.\ Grav.\  {\bf 30}, 104002 (2013)
  [arXiv:1204.1057 [hep-th]].

\bibitem{Dolan:2001ih} 
  L.~Dolan, C.~R.~Nappi and E.~Witten,
  ``Conformal operators for partially massless states,''
  JHEP {\bf 0110}, 016 (2001)
  [hep-th/0109096].
  
\bibitem{Deser:2003gw} 
  S.~Deser and A.~Waldron,
  ``Arbitrary spin representations in de Sitter from dS / CFT with applications to dS supergravity,''
  Nucl.\ Phys.\ B {\bf 662}, 379 (2003)
  [hep-th/0301068].
  
\bibitem{Bekaert:2013zya} 
  X.~Bekaert and M.~Grigoriev,
  ``Higher order singletons, partially massless fields and their boundary values in the ambient approach,''
  Nucl.\ Phys.\ B {\bf 876}, 667 (2013)
  [arXiv:1305.0162 [hep-th]].
  
\bibitem{Didenko:2012tv} 
  V.~E.~Didenko and E.~D.~Skvortsov,
  ``Exact higher-spin symmetry in CFT: all correlators in unbroken Vasiliev theory,''
  JHEP {\bf 1304}, 158 (2013)
  [arXiv:1210.7963 [hep-th]].

\bibitem{Freedman:1998tz} 
  D.~Z.~Freedman, S.~D.~Mathur, A.~Matusis and L.~Rastelli,
  ``Correlation functions in the CFT(d) / AdS(d+1) correspondence,''
  Nucl.\ Phys.\ B {\bf 546}, 96 (1999)
  [hep-th/9804058].
  
\bibitem{Arutyunov:1999nw} 
  G.~Arutyunov and S.~Frolov,
  ``Three point Green function of the stress energy tensor in the AdS / CFT correspondence,''
  Phys.\ Rev.\ D {\bf 60}, 026004 (1999)
  [hep-th/9901121].
  
\bibitem{Vasiliev:2012vf} 
  M.~A.~Vasiliev,
  ``Holography, Unfolding and Higher-Spin Theory,''
  J.\ Phys.\ A {\bf 46}, 214013 (2013)
  [arXiv:1203.5554 [hep-th]].
  
\bibitem{Giombi:2009wh} 
  S.~Giombi and X.~Yin,
  ``Higher Spin Gauge Theory and Holography: The Three-Point Functions,''
  JHEP {\bf 1009}, 115 (2010)
  [arXiv:0912.3462 [hep-th]].

\bibitem{Giombi:2010vg} 
  S.~Giombi and X.~Yin,
  ``Higher Spins in AdS and Twistorial Holography,''
  JHEP {\bf 1104}, 086 (2011)
  [arXiv:1004.3736 [hep-th]].
  
\bibitem{Colombo:2012jx} 
  N.~Colombo and P.~Sundell,
  ``Higher Spin Gravity Amplitudes From Zero-form Charges,''
  arXiv:1208.3880 [hep-th].

\bibitem{Didenko:2013bj} 
  V.~E.~Didenko, J.~Mei and E.~D.~Skvortsov,
  ``Exact higher-spin symmetry in CFT: free fermion correlators from Vasiliev Theory,''
  Phys.\ Rev.\ D {\bf 88}, 046011 (2013)
  [arXiv:1301.4166 [hep-th]].

\bibitem{Wigner} 
  W.~B.~Case,
  ``Wigner functions and Weyl transforms for pedestrians,''
  Am.\ J.\ Phys.\  {\bf 76}, 937 (2008).

\bibitem{Goheer:2002vf} 
  N.~Goheer, M.~Kleban and L.~Susskind,
  ``The Trouble with de Sitter space,''
  JHEP {\bf 0307}, 056 (2003)
  [hep-th/0212209].
  
\bibitem{Bars:2014vca} 
  I.~Bars and D.~Rychkov,
  ``Background Independent String Field Theory,''
  arXiv:1407.4699 [hep-th].
  
\bibitem{Bars:2014jca} 
  I.~Bars and D.~Rychkov,
  ``Is String Interaction the Origin of Quantum Mechanics?,''
  Phys.\ Lett.\ B {\bf 739}, 451 (2014)
  [arXiv:1407.6833 [hep-th]].
  
\bibitem{Halpern:2015cia} 
  I.~Halpern and Y.~Neiman,
  ``Quantum fields and entanglement on a curved lightfront,''
  arXiv:1502.04106 [hep-th].
  
\bibitem{Bunch:1978yq} 
  T.~S.~Bunch and P.~C.~W.~Davies,
  ``Quantum Field Theory in de Sitter Space: Renormalization by Point Splitting,''
  Proc.\ Roy.\ Soc.\ Lond.\ A {\bf 360}, 117 (1978).

\bibitem{Leutwyler:1970wn} 
  H.~Leutwyler, J.~R.~Klauder and L.~Streit,
  ``Quantum field theory on lightlike slabs,''
  Nuovo Cim.\ A {\bf 66}, 536 (1970).
  
\bibitem{Allen:1985ux} 
  B.~Allen,
  ``Vacuum States in de Sitter Space,''
  Phys.\ Rev.\ D {\bf 32}, 3136 (1985).
  
\bibitem{Maskawa:1975ky} 
  T.~Maskawa and K.~Yamawaki,
  ``The Problem of P+ = 0 Mode in the Null Plane Field Theory and Dirac's Method of Quantization,''
  Prog.\ Theor.\ Phys.\  {\bf 56}, 270 (1976).

\bibitem{Yamawaki:1997cj} 
  K.~Yamawaki,
  ``Zero mode and symmetry breaking on the light front,''
  In *Les Houches 1997, New non-perturbative methods and quantization on the light cone* 301-309.
  [hep-th/9707141].
  
\bibitem{Balasubramanian:2002zh} 
  V.~Balasubramanian, J.~de Boer and D.~Minic,
  ``Notes on de Sitter space and holography,''
  Class.\ Quant.\ Grav.\  {\bf 19}, 5655 (2002)
  [Annals Phys.\  {\bf 303}, 59 (2003)]
  [hep-th/0207245].
  
\bibitem{Neiman:2015wma} 
  Y.~Neiman,
  ``Higher-spin gravity as a theory on a fixed (anti) de Sitter background,''
  JHEP {\bf 1504}, 144 (2015)
  [arXiv:1502.06685 [hep-th]].
  
\bibitem{Gao:2000ga} 
  S.~Gao and R.~M.~Wald,
  ``Theorems on gravitational time delay and related issues,''
  Class.\ Quant.\ Grav.\  {\bf 17}, 4999 (2000)
  [gr-qc/0007021].
  
\bibitem{Leblond:2002ns} 
  F.~Leblond, D.~Marolf and R.~C.~Myers,
  ``Tall tales from de Sitter space 1: Renormalization group flows,''
  JHEP {\bf 0206}, 052 (2002)
  [hep-th/0202094].
  
\end{thebibliography}
\end{document}